\pdfoutput=1
\documentclass[letterpaper,twocolumn,10pt]{article}
\usepackage{amssymb}
\usepackage{usenix-2020-09}
\usepackage{tikz}
\usetikzlibrary{arrows.meta,positioning,fit,backgrounds,calc,decorations.pathreplacing}
\usepackage{xcolor,xspace}
\usepackage{xurl}
\hypersetup{breaklinks=true}
\usepackage{epsfig,graphicx,subfig,amsmath} %
\usepackage{endnotes,multirow,algorithm,algpseudocode}
\usepackage{enumitem}
\usepackage[T1]{fontenc}
\usepackage{color,soul}
\usepackage{booktabs}
\usepackage{colortbl}
\usepackage{pifont}
\newcommand{\cmark}{\textcolor{green!60!black}{\ding{51}}}%
\newcommand{\xmark}{\textcolor{red!80!black}{\ding{55}}}%
\newcommand{\tmark}{\textcolor{red!80!black}{\ding{115}}}%
\usepackage{balance}
\providecommand{\figfontS}{\fontfamily{Crlt-LF}\fontsize{8}{9.2}\selectfont}
\definecolor{kBlueFill}{HTML}{DEEBF7}
\definecolor{kBlueLine}{HTML}{41719C}
\definecolor{kGreenPale}{HTML}{E2F0D9}
\definecolor{kGreenFill}{HTML}{C5E0B4}
\definecolor{kGreenLow}{HTML}{92D050}
\definecolor{kGreenLine}{HTML}{548235}
\definecolor{kGreenDark}{HTML}{385723}
\definecolor{kTealFill}{HTML}{CBE7E6}
\definecolor{kTealLine}{HTML}{3D8380}
\definecolor{kLavFill}{HTML}{EBEBFF}
\definecolor{kLavLine}{HTML}{6763B5}
\definecolor{kPeach}{HTML}{FBE5D6}
\definecolor{kPeachLine}{HTML}{C55A11}
\definecolor{kYellow}{HTML}{FFF2CC}
\definecolor{kGold}{HTML}{BF9000}
\definecolor{kGrayFill}{HTML}{E7E6E6}
\definecolor{kGrayLine}{HTML}{7F7F7F}
\definecolor{kInk}{HTML}{000000}
\tikzset{
  kfig/.style={font=\figfontS, line width=0.6pt, >={Latex[length=1.6mm,width=1.4mm]}},
  kcell/.style={draw, minimum width=#1, minimum height=3.2mm, inner sep=0pt, outer sep=0pt, anchor=south west},
  kbox/.style={draw, rounded corners=1.5pt, align=center, inner sep=2pt},
  kblock/.style={-{Triangle[length=1.6mm,width=2.6mm]}, line width=1.6pt},
  karrow/.style={->, line width=0.7pt},
}

\newcommand{\sys}{KatoANN\xspace}

\newcommand{\cabana}{CABANA\xspace}
\newcommand{\scann}{ScaNN\xspace}
\newcommand{\faiss}{Faiss\xspace}

\hypersetup{breaklinks=true}
\begin{document}

\title{\sys: Enabling Near-Exact Vector Search at the Speed of Lossy Scans with CPU Matrix Engines}
\author{
{\rm Soobin Cho and Gyuyeong Kim}\\
Yonsei University
}

\maketitle

\begin{abstract}
Emerging CPU matrix engines, such as Intel AMX, enable high throughput by batching multiple queries.
However, they support only low-precision formats, which can significantly degrade recall (i.e., search quality).
We present \sys{}, the first approximate nearest neighbor (ANN) search system that achieves both high throughput and high recall simultaneously on CPU matrix engines.
Specifically, \sys represents each vector using a 16-bit code, splits it into high- and low-byte planes, and computes near-exact distances using 8-bit matrix multiplications that the matrix engine supports.
To reduce extra multiplication overhead, it ranks all candidates by their high bytes alone and refines only a few survivors with the full distance.
\sys also reduces memory footprint since it does not maintain a full-precision copy of the vectors.
We implement a \sys{} prototype on Intel AMX.
Our experimental results show that \sys{} outperforms existing partition-based systems across the high-recall range.

\end{abstract}

\section{Introduction}
\label{introduction}

Approximate nearest neighbor (ANN) search over high-dimensional embeddings is a core building block of modern AI applications, such as recommendation~\cite{covington2016youtube,naumov2019dlrm} and retrieval-augmented generation (RAG)~\cite{lewis2020rag}.
Exact $K$-nearest neighbor (KNN) search returns the true neighbors with a recall of 1, but it must scan the entire dataset.
ANN search avoids this expensive scan, and an effective ANN system should achieve recall close to 1 (e.g., 0.99).
This is because, for some applications (e.g., RAG for law and medicine~\cite{clerc,medrag}), a missed neighbor can drop critical evidence that the large language model (LLM) needs.
Agentic RAG also issues many searches per task~\cite{ircot,react,searchr1} and may run subagents that search in parallel~\cite{anthropic-multiagent}.
Under such high concurrency, ANN search can become a performance bottleneck, leading to excessive end-to-end latency~\cite {rago,piperag}.

\begin{figure}[t]
\centering
\includegraphics[width=\linewidth]{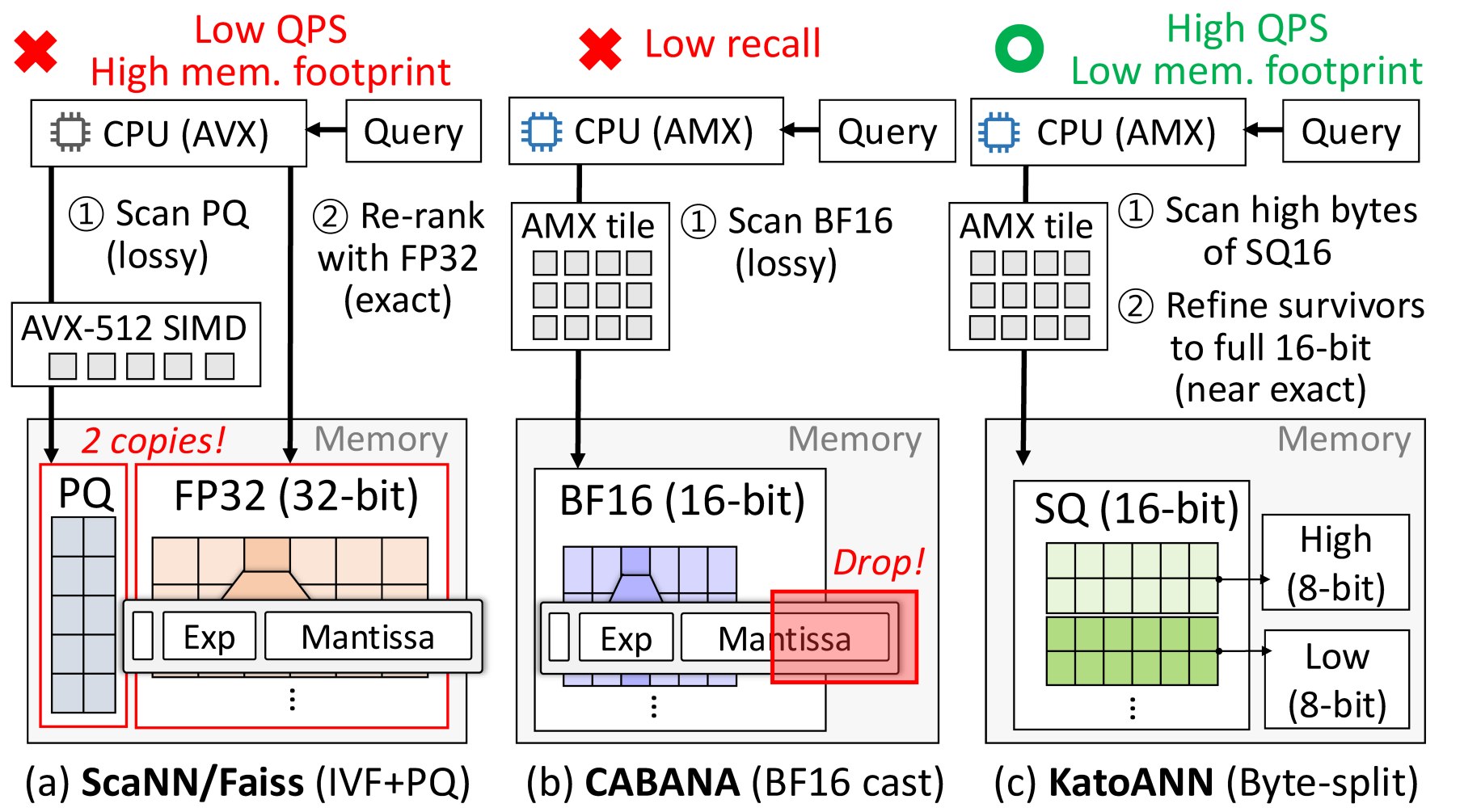}
\caption{\sys{} provides high throughput and high recall with a small memory footprint through an adaptive rank-then-refine scan over byte-split codes on a CPU matrix engine.}
\label{fig:arch-compare}
\vspace{-3mm}
\end{figure}

Partition-based search~\cite{milvus,faiss,scann,spann} is one of the two dominant ANN index families, alongside graph-based search~\cite{diskann,pipeann}.
It spends most of its query time scanning the vectors in the selected partitions.
This distance scan is typically \emph{memory-bandwidth-bound}, and its throughput depends on the number of bytes read per candidate vector.
To read fewer bytes, these systems typically scan compressed codes instead of raw vectors, such as product quantization (PQ)~\cite{pq} codes.
These low-precision codes reduce memory traffic but can limit recall.
To achieve high recall, \scann~\cite{scann} and \faiss~\cite{faiss} use two passes, a coarse pass over the compressed codes followed by a re-ranking pass over full-precision (FP32) vectors.
However, re-ranking requires retaining a full-precision copy of the vectors and extra computation, reducing both memory efficiency and throughput, as illustrated in Figure~\ref{fig:arch-compare}~(a).

Meanwhile, recent CPUs include built-in matrix engines (e.g., Intel Advanced Matrix Extensions (AMX)~\cite{amx}) that can multiply two blocks of numbers in a single instruction.
These engines can provide high search throughput by batching multiple queries.
For example, \cabana~\cite{cabana} is a recent work that batches queries that probe the same cluster so that one matrix multiplication scans the cluster for all of them.
However, matrix engines reach this throughput only with low-precision data types such as the 16-bit floating-point format BF16 and 8-bit integers (INT8).
Therefore, \cabana scans lossy BF16 codes and often fails to achieve high recall (Figure~\ref{fig:arch-compare}~(b)).
Re-ranking with a full-precision copy, as with \faiss, would help, but it would reintroduce the memory cost and computation overhead.
This leads us to ask: \emph{Can an ANN search achieve both high throughput and high recall on CPU matrix engines without storing a second copy of the vectors?}

To answer this question affirmatively, we present \sys, the first ANN search system to achieve both high recall and high throughput on CPU matrix engines while maintaining memory efficiency.
Our key insight is that the precision of vector distances need not be limited by the precision of hardware-supported data types.
Specifically, the inner product of two 16-bit integer codes can be reconstructed exactly from four 8-bit integer inner products computed natively by the engine.
Based on this insight, \sys stores each vector only once as a near-exact 16-bit scalar-quantized (SQ16) code and splits each coordinate into an 8-bit high byte and an 8-bit low byte, a technique we call the \emph{byte-split} (\S\ref{sec:design-precision}).
This enables \sys to compute near-exact SQ16-precision distances without maintaining a second copy of the vectors.

However, the byte-split requires four multiplications per candidate instead of one.
This cost matters because batching lets \sys{} process many queries that probe the same cluster together, reusing each vector fetched from memory across these queries.
As a result, the distance scan becomes \emph{compute-bound}, with throughput determined by the number of multiply-add operations rather than the bytes read per candidate.
To sustain high throughput, \sys{} employs an \emph{adaptive rank-then-refine scan} (\S\ref{sec:design-cascade}), which first ranks all candidates using their high bytes with a single INT8 multiplication and then computes full SQ16 distances only for a small number of survivors.
For high-dimensional embeddings, the ranking stage further reduces memory traffic by reading only a short prefix of each vector.
Finally, \sys{} organizes codes in the order consumed by the matrix engine and prevents centroid scanning and top-$K$ selection from stalling the faster distance scan (\S\ref{sec:design-busy}).

We implement a \sys{} prototype on a commodity server with Intel AMX.
Our experimental results show that \sys{} is the fastest partition-based system across the high-recall range and matches or outperforms graph-based systems.
For example, on 3,072-dimensional embeddings at recall@$10 \ge 0.99$, \sys{} has higher throughput than \cabana by $3.13\times$.

In summary, this work makes the following contributions:
\begin{itemize}[noitemsep,leftmargin=*]
\item We show that ANN search on CPU matrix engines can achieve both high recall and high throughput through our byte-split design.
\item We propose \sys{}, to our knowledge the first high-throughput, high-recall ANN search system on CPU matrix engines.
\item We implement a \sys{} prototype on a commodity Intel AMX server and evaluate it on diverse datasets.
\end{itemize}

\section{Background and Motivation \label{motivation}}
\subsection{The Trade-off in the IVF Distance Scan
\label{sec:motiv-wall}}

\begin{figure}[t]
\centering
\includegraphics[width=\linewidth]{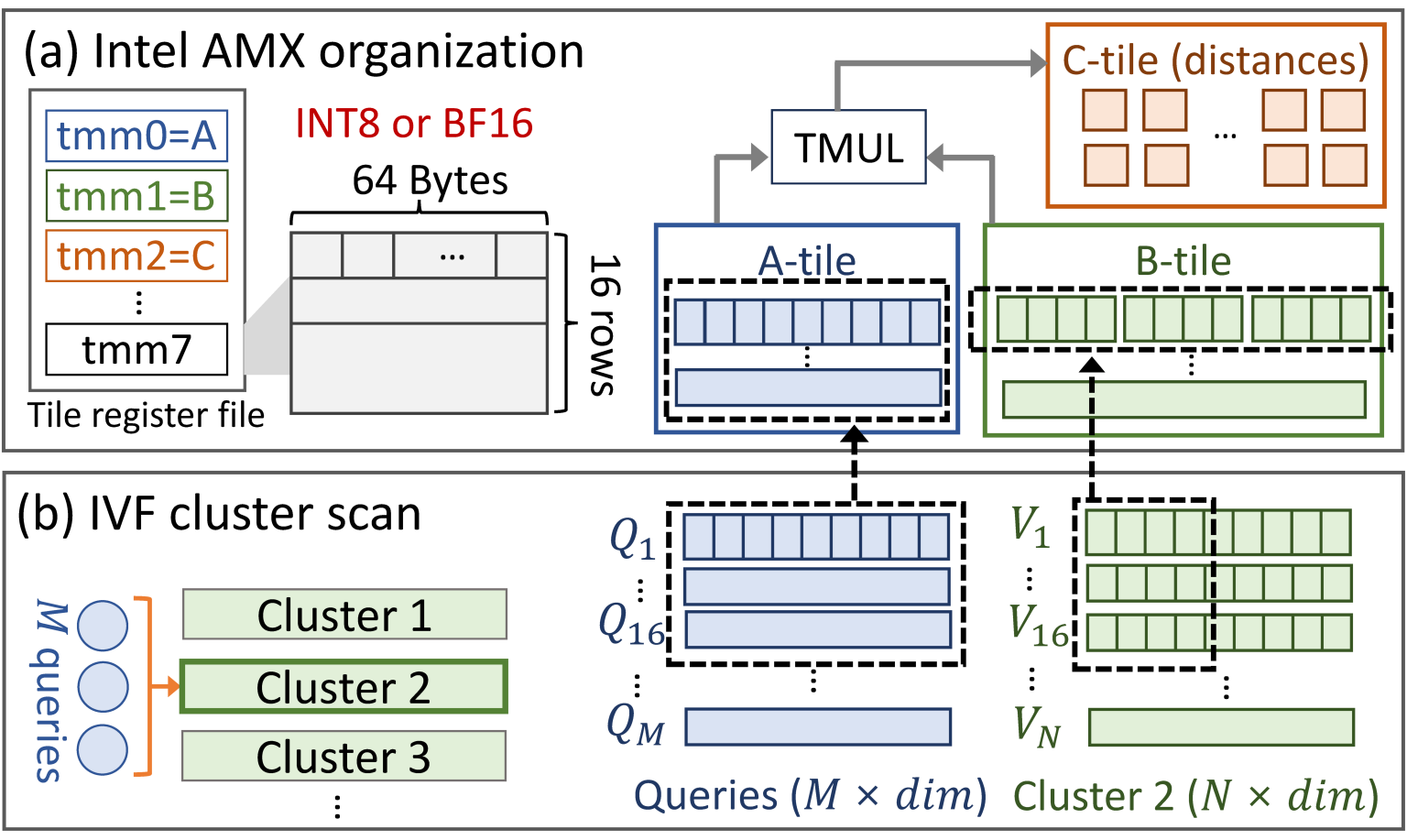}
\caption{
AMX and a batched IVF scan.
\emph{
(a)~AMX provides two-dimensional tile registers and tile-multiply instructions.
(b)~Queries grouped by probed cluster form one input matrix, and the vector codes in that cluster form the other.
}
\label{fig:amx-scan}}
\vspace{-3mm}
\end{figure}

\noindent\textbf{IVF search scans.}
Unlike graph indexes that traverse a proximity graph and compute distances with irregular memory access~\cite{hnsw,diskann,pipeann}, an inverted-file (IVF) index partitions vectors into clusters and scans the lists of the clusters closest to the query~\cite{faiss,scann,spann}, making distance computations easy to batch.
Insertions are also straightforward, as each new vector is appended to its assigned list.

\noindent\textbf{A scan is memory-bandwidth-bound.}
The speed $T$ of the distance scan follows the roofline model,
\begin{equation} T \;\approx\; \min\!\big(\pi,\; I\,\beta\big), \qquad I \;=\; \frac{\#\,\text{mul-add ops}} {\#\,\text{bytes read}}, \label{eq:roofline} \end{equation}
where $\pi$ is the peak arithmetic rate, $\beta$ is the memory bandwidth, and $I$ is the arithmetic intensity of the scan, the number of multiply-add operations per byte read from memory.
In a conventional single-query scan, each loaded code serves only one distance computation, and $I$ stays at $O(1)$ multiply-add operations per byte.
As a result, the scan runs at $I\beta$, well below the compute peak $\pi$.

\noindent\textbf{Compact codes sacrifice recall.}
To read fewer bytes, IVF systems scan compact codes, such as the PQ codes of ScaNN~\cite{scann} and Faiss~\cite{faiss} and the scalar quantization (SQ) codes of Faiss.
These codes may not distinguish nearby coordinate values.
At high recall, probing more clusters may stop helping, because quantization error becomes the limiting factor.

\noindent\textbf{Re-ranking adds memory and compute costs.}
Re-ranking the survivors with full-precision vectors recovers recall, but it stores a full-precision copy alongside the compact codes and adds a second distance pass.
Other recent approaches keep accurate distances but stop a comparison once enough dimensions have been examined~\cite{adsampling,dade}.
Their candidate-dependent and branch-heavy execution does not map well to dense batched computation.

\subsection{Opportunity with CPU Matrix Engines
\label{sec:motiv-amx}}
\begin{figure}[t]
\centering
\subfloat[Bottleneck shift with batching]{\includegraphics[width=0.48\linewidth]{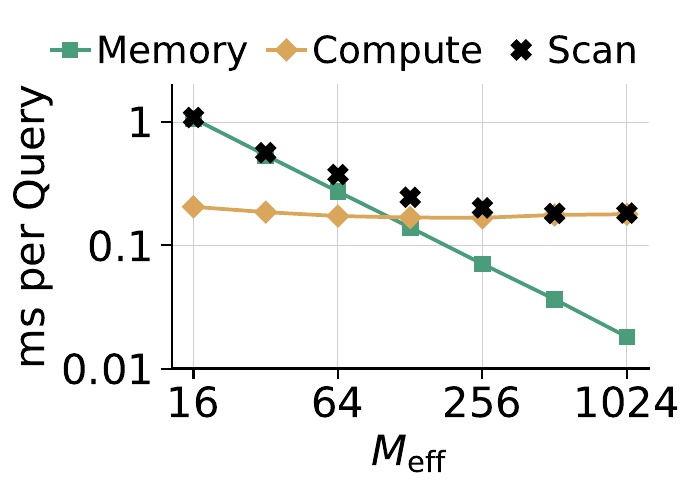}}\hfill
\subfloat[Recall ceiling of \cabana]{\includegraphics[width=0.48\linewidth]{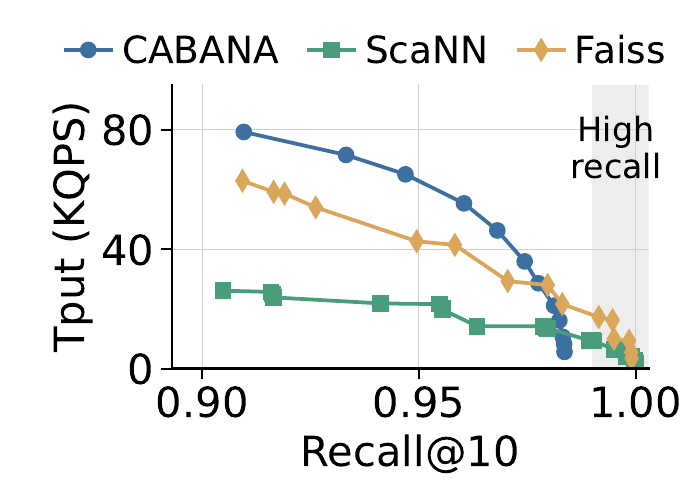}}
\caption{
Batching makes the scan compute-bound, and \cabana~\cite{cabana}, an existing batching system, limits recall (search quality). $M_{\mathrm{eff}}$: \# of queries that probe the same list. 
\label{fig:motiv-amx}}
\vspace{-3mm}
\end{figure}

\noindent\textbf{Matrix engines match batched distance scans.}
Existing IVF systems run the scan on SIMD units, which compute multiply-add operations over one-dimensional vector registers.
A matrix engine in a commodity CPU, such as Intel AMX~\cite{amx} or the Arm Scalable Matrix Extension (SME)~\cite{arm-sme}, can execute far more operations per instruction, because it multiplies one two-dimensional block against another.
Figure~\ref{fig:amx-scan} shows how a batched IVF scan maps onto AMX.
AMX multiplies a block of queries in \emph{tile} registers, the A-tile, by a block of vector codes, the B-tile, and accumulates a dense result block.
For each probed cluster, the vector codes of the cluster form the B-tiles, and the queries that probe the cluster form the A-tiles.

\noindent\textbf{Batching makes the scan compute-bound.}
In a single-query scan, each loaded vector code serves only one query.
With cluster-aware grouping~\cite{cabana}, the same block instead serves all queries in the batch that probe the same list.
We call the number of these queries the \emph{effective batch size} $M_{\mathrm{eff}}$.
It counts only the queries that share a list, not all queries in the batch.
The arithmetic intensity then rises from $I \approx 1$ to $I \approx M_{\mathrm{eff}}$, because the same bytes are reused across many queries.
By Eq.~\ref{eq:roofline}, the scan becomes \emph{compute-bound} once
\begin{equation}
M_{\mathrm{eff}} \;\gtrsim\; M^\star
\;=\; \frac{\pi}{\beta}.
\label{eq:crossover}
\end{equation}
Figure~\ref{fig:motiv-amx}~(a) measures this transition with a BF16 distance kernel ($d{=}1024$, $2$M vectors, $28$ threads).
Memory estimates the transfer time from measured DRAM traffic and bandwidth, and Compute scales the cache-resident runtime of the same kernel to $2$M vectors.
As $M_{\mathrm{eff}}$ increases, the measured scan time moves from Memory toward Compute.
Therefore, cross-query reuse makes computation the dominant cost, and a matrix engine reduces this cost.
With the same kernel, AMX has higher throughput than AVX-512 by $2.6\times$ at $M_{\mathrm{eff}}{=}16$, and the gap grows to $14.8\times$ at $M_{\mathrm{eff}}{=}1{,}024$.

\begin{figure*}[t]
\centering
\includegraphics[width=0.83\textwidth]{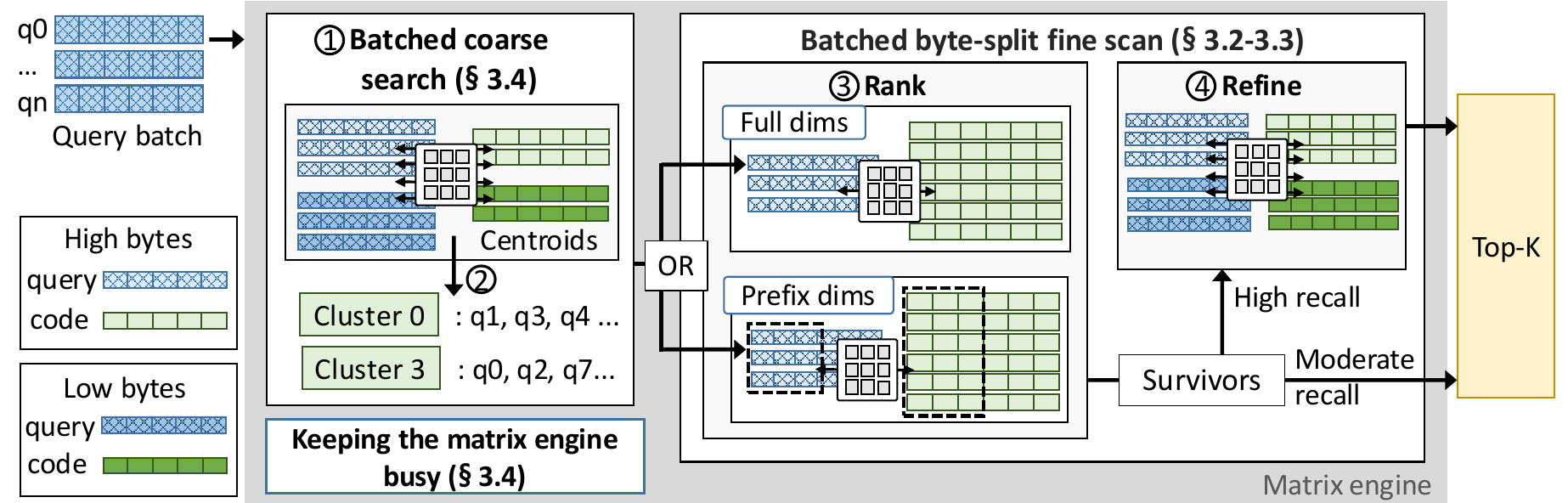}
\caption{
The \sys{} pipeline.
\emph{
(1) A batched coarse search scores the batch against all centroids and chooses the clusters to probe.
(2) The fine scan groups queries by probed cluster.
(3) It ranks all probed candidates with a cheap high-byte key, which reads only a rotated prefix of the high-byte plane on high-dimensional data, after the batch is rotated once.
(4) When the recall target requires higher precision, it refines only the survivors with the full SQ16 byte-split distance.
}
\label{fig:arch}}
\vspace{-3mm}
\end{figure*}

\noindent\textbf{Matrix engines are fast only on low-precision formats.}
However, matrix engines reach their high throughput only with low-precision operands.
Intel AMX, for example, multiplies only INT8 and 16-bit floating-point tiles such as BF16.
\cabana~\cite{cabana} maps the cluster-aware IVF fine scan to efficient matrix multiplication, and it stores lossy BF16 codes because the engine supports that format directly.
BF16 spends eight of its 16 bits on the exponent and keeps only eight significant bits.
For example, 1.000 and 1.003 become the same BF16 value.
Candidates with similar distances can then swap places in the ranking, and \cabana can miss true neighbors at high recall targets.
Figure~\ref{fig:motiv-amx}~(b) compares \cabana with \scann and \faiss on Cohere-768-1M (\S\ref{evaluation}).
\cabana has the highest throughput up to a recall of about $0.98$, but it cannot reach $0.99$.
\scann and \faiss, which re-rank with full-precision vectors, reach $0.99$ at much lower throughput.
A scan over full-precision codes would keep recall, but it would give up the speed of low-precision operands.
This implies that naively leveraging a CPU matrix engine is not enough.

\subsection{Design Rationale
\label{sec:motiv-approach}}
\begin{table}[t]
\centering
\small
\caption{
Comparison to existing approaches.
\label{tab:comparison}}
\setlength{\tabcolsep}{2pt}
\renewcommand{\arraystretch}{1.15}
\resizebox{\columnwidth}{!}{
\begin{tabular}{l c c c c c}
\hline
 & \shortstack{\cabana~\cite{cabana}}
 & \shortstack{FastScan\\\cite{faiss}}
 & \shortstack{ADS/DADE\\\cite{adsampling,dade}}
 & \shortstack{\scann~\cite{scann}}
 & \sys{} \\
\hline
Resident data
 & BF16
 & PQ4${+}$FP32
 & FP32/projection
 & lossy${+}$FP32
 & SQ16 \\
High recall
 & \xmark
 & \tmark
 & \cmark
 & \cmark
 & \cmark \\
High throughput
 & \tmark
 & \tmark
 & \xmark
 & \xmark
 & \cmark \\
Memory efficiency
 & \cmark
 & \xmark
 & \xmark
 & \xmark
 & \cmark \\
\hline
\end{tabular}
}
\par\smallskip
\vspace{-3mm}
\end{table}
To achieve high recall on CPU matrix engines that provide only low-precision operands, \sys{} computes exact 16-bit distances by combining several 8-bit products, in the same way as long multiplication.
For example, $12 \times 34$ requires only the single-digit products $1{\times}3$, $1{\times}4$, $2{\times}3$, and $2{\times}4$, each shifted by its place value before they are added.
Specifically, \sys{} stores each coordinate as a 16-bit code (SQ16) and treats its high byte and low byte as two digits.
The four products of their bytes then run as INT8 operations on the matrix engine, and shifting and adding them gives the exact SQ16 dot product.
SQ16 takes as much memory as BF16, but it has 16 bits of resolution instead of BF16's eight significant bits.
Therefore, SQ16 can keep small differences that BF16 rounds away.

However, this exactness costs four multiplications per candidate instead of one, and batching has already made multiplication the bottleneck of the scan.
One observation is that the high byte is the leading digit of each code, so the high-byte product alone yields a rough distance with a single multiplication rather than four.
Therefore, \sys{} ranks all candidates by the high-byte product and computes all four products only for candidates that could still enter the top-$K$.
With this, most candidates cost no more than in a conventional low-precision scan.
In addition, both distances come from the same stored code, so \sys{} keeps no second copy of the vectors.

\noindent\textbf{Comparison with existing approaches.}
As Table~\ref{tab:comparison} shows, each existing approach falls short in recall, throughput, or memory efficiency, whereas \sys{} provides all three.

\section{\sys{} Design \label{design}}

\subsection{System Overview \label{sec:design-overview}}

\sys{} is a partition-based ANN search system that stores each database vector only once, as a resident SQ16 code, and derives every distance from that code.
Compared with a compact scan code plus a full-precision copy for re-ranking, this design needs no second copy, and its refinement reads the same code instead of the full-precision vectors.
\sys consists of three parts as follows:
\begin{itemize}[noitemsep,leftmargin=*]
\item The \textbf{byte-split distance} (\S\ref{sec:design-precision}) computes SQ16 distances from four INT8 matrix multiplications.
\item The \textbf{adaptive rank-then-refine scan} (\S\ref{sec:design-cascade}) ranks all candidates by their high bytes, on a rotated prefix for high-dimensional data, and computes the full distance only for the survivors.
\item The \textbf{batched scan pipeline} (\S\ref{sec:design-busy}) groups the queries of a batch by probed cluster, stores the codes in the order in which the engine loads them, and runs the coarse search and top-$K$ selection without stalling the engine.
\end{itemize}

\noindent\textbf{Life of a query batch.}
Figure~\ref{fig:arch} shows how a batch moves through \sys{}.
For high-dimensional data, \sys{} rotates the batch once into the basis of the dimension cascade (\S\ref{sec:design-dimc}) before the search.
A serving batch first enters the coarse search, which scores the batch against all centroids and selects the $\text{nprobe}$ nearest clusters.
\sys{} then inverts the schedule from query order to cluster order.
Each cluster is scanned once, in one dense matrix multiply against all the queries that probe it.
The fine scan orders the probed candidates by the high-byte key.
At lower recall targets, this already yields the final top-$K$.
At higher ones, the best $S$ candidates per query, which we call \emph{survivors}, continue into the full SQ16 refinement.
The number of survivors is $S=R\!\cdot\!K$, where $K$ is the number of returned neighbors and $R$ is the survivor factor.
For high-dimensional data, a one-time calibration chooses a prefix length for each index.

\subsection{Byte-Split Distance
\label{sec:design-precision}}

\sys{} uses a uniform SQ16 quantization grid for both queries and database vectors, with an offset defining the lower bound and a step size specifying the interval between quantization levels.
Each SQ16 coordinate of a database vector is split into its high byte and its low byte.
\sys{} stores the high bytes of a cluster in one plane and the low bytes in another, and these two planes are the only copy of the vectors.
The resulting distance is not bit-identical to FP32, but it is near-exact for ANN ranking at the recall targets we consider.

\noindent\textbf{Four INT8 products.}
We write an SQ16 coordinate as $x = 2^8 x_h + x_l$ and $q = 2^8 q_h + q_l$.
The dot product then decomposes as
\begin{equation}
\langle q, x\rangle
=
2^{16}\langle q_h, x_h\rangle
+
2^8\big(\langle q_h, x_l\rangle + \langle q_l, x_h\rangle\big)
+
\langle q_l, x_l\rangle .
\label{eq:bytesplit}
\end{equation}
Each term is an INT8 matrix multiplication over byte planes.
The high and low bytes of a block of queries fill two A-tiles, the high and low bytes of a block of database codes fill two B-tiles, and four tile multiplications produce the four terms for the whole block.
The shifted sum then reconstructs the full SQ16 dot product.

\noindent\textbf{Distances and exactness.}
For the squared Euclidean ($L_2$) distance, \sys{} combines this dot product with the squared norm of each database code, precomputed and stored with the code, and the squared norm of each query, computed once per batch:
\begin{equation}
\lVert q-x\rVert^2
=
\lVert q\rVert^2 + \lVert x\rVert^2 - 2\langle q,x\rangle .
\label{eq:l2-dot}
\end{equation}
Because the queries and the database vectors share one SQ16 grid, the offset of the grid cancels in their difference, and \sys{} multiplies the integer distance by the squared step size of the grid once at the end.
Because the two bytes represent the SQ16 code exactly, the four partial products reconstruct the SQ16 dot product with no rounding beyond SQ16 quantization.
Each INT8 tile multiply accumulates into 32-bit integers, and \sys{} converts the partial sums to 64-bit integers before the $2^8$ and $2^{16}$ shifts, so the sum does not overflow for the dimensions we target.
For datasets whose vectors are already 8-bit unsigned integers (UINT8), \sys{} multiplies these bytes directly.
For cosine distance, it uses the same path on normalized vectors.

\begin{figure}[t]
\centering
\includegraphics[width=\columnwidth]{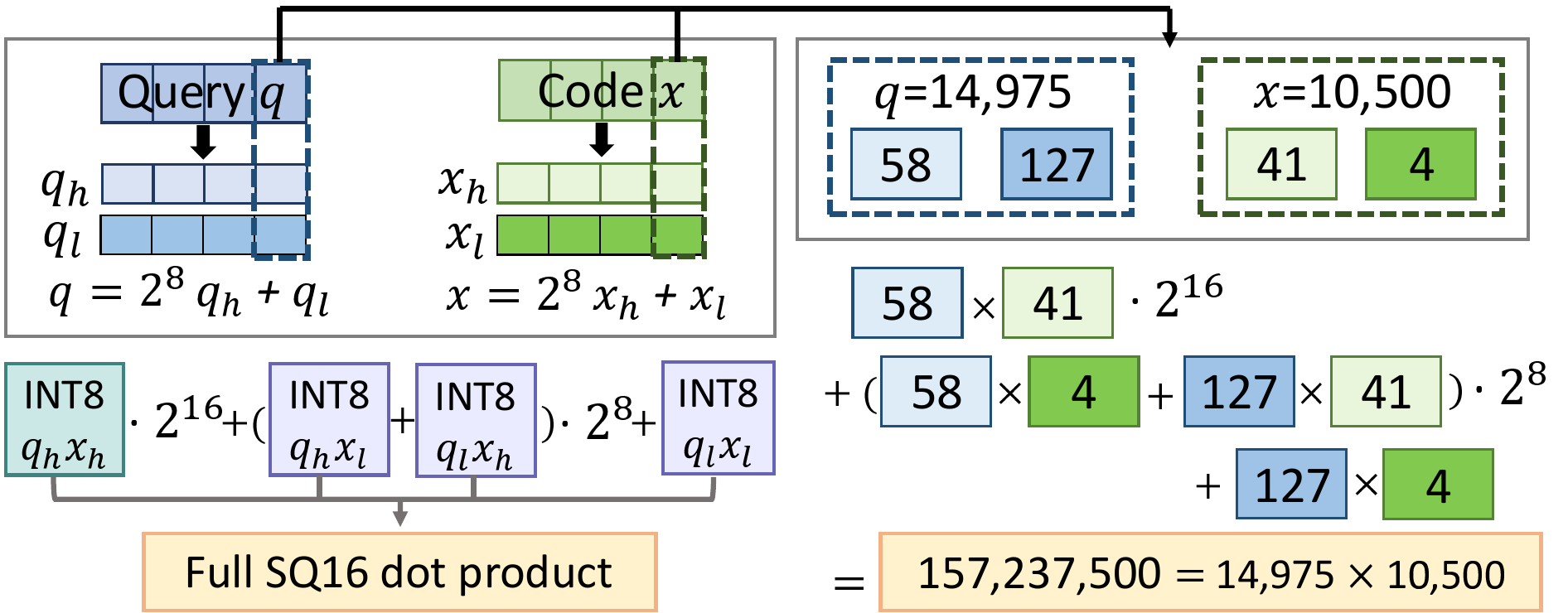}
\caption{An example of the byte-split distance.\label{fig:bytesplit-ex}}
\end{figure}

Figure~\ref{fig:bytesplit-ex} shows the byte-split distance for one coordinate.
Consider one coordinate with the query value $q = 14{,}975$ and the database value $x = 10{,}500$.
Their bytes are $q_h = 58$, $q_l = 127$, $x_h = 41$, and $x_l = 4$, because $14{,}975 = 58 \cdot 256 + 127$ and $10{,}500 = 41 \cdot 256 + 4$.
The four byte products are $58 \cdot 41 = 2{,}378$, $58 \cdot 4 = 232$, $127 \cdot 41 = 5{,}207$, and $127 \cdot 4 = 508$.
Weighted by $2^{16}$, $2^{8}$, $2^{8}$, and $1$, they sum to $157{,}237{,}500$, which equals $14{,}975 \times 10{,}500$ exactly.
Summing these values over all coordinates gives the dot product of the query and the vector.

\subsection{Adaptive Rank-then-Refine Scan
\label{sec:design-cascade}
\label{sec:design-dimc}}

Our full byte-split distance needs four INT8 multiplications per candidate, whereas a conventional low-precision scan needs only one multiplication.
This extra multiplication overhead can degrade throughput.
For high-dimensional data, even a single product per candidate can still dominate the scan because it reads every coordinate.
To address these problems, \sys{} ranks all candidates with the cheapest of the four products and refines only a few survivors with the full distance.
We call this scan \emph{rank-then-refine}.

On high-dimensional data, it ranks on a short prefix of rotated coordinates, which we call the \emph{dimension cascade}.
The dimension cascade is thus rank-then-refine on a prefix, with a cheap pass over all candidates and the full distance for the survivors.

\begin{figure}[t]
\centering
\includegraphics[width=\columnwidth]{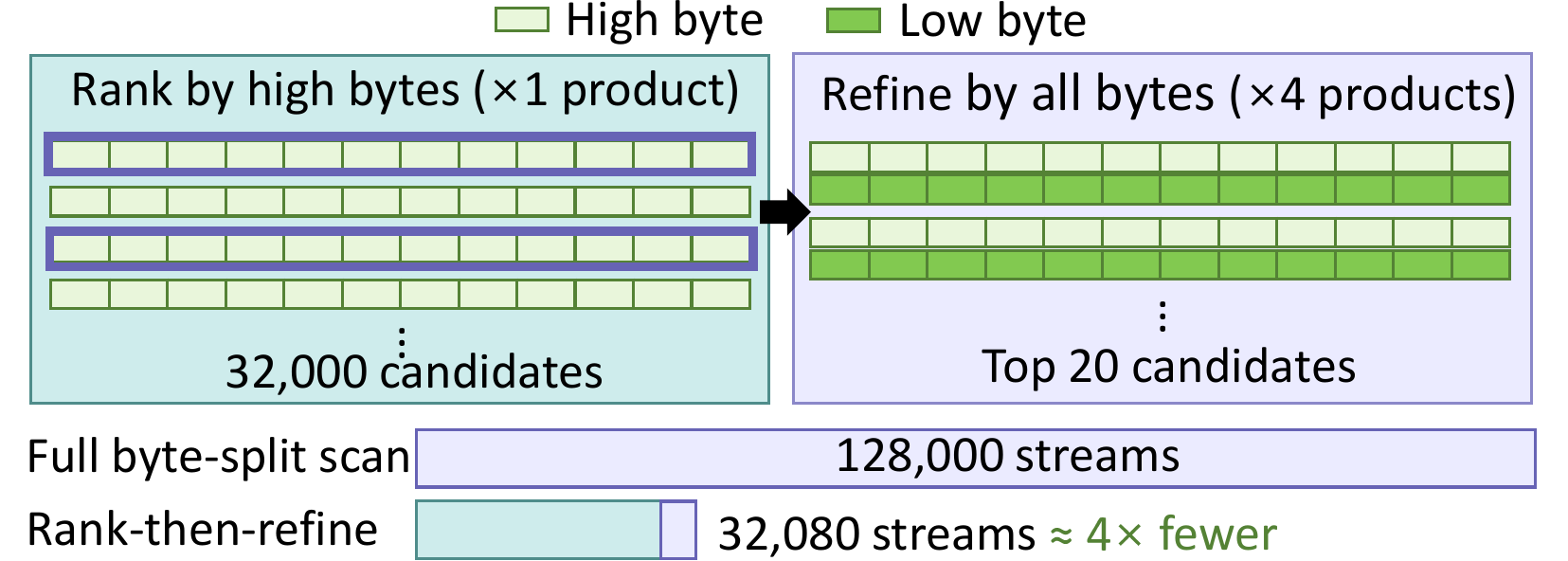}
\caption{An example of the rank-then-refine scan.\label{fig:cascade-ex}}
\end{figure}

\noindent\textbf{Rank all, refine few.}
Most probed candidates are far from the top-$K$, and a coarse distance is enough to exclude them.
The byte split exposes such a distance as a cheap ranking key.
The high-by-high term $\langle q_h,x_h\rangle$ is one of the four partial products.
Together with the precomputed squared norms of the high bytes, it gives the exact $L_2$ distance between the high-byte truncations of the query and the database vector.
This key reads only the high-byte plane.
\sys{} uses this key in two phases.
Phase~1 ranks all probed candidates with the high-byte key, which takes one INT8 partial product per candidate.
If this ranking is enough for the target recall, \sys{} returns its top-$K$ directly.
Otherwise, Phase~2 takes the top $S=R\!\cdot\!K$ candidates of the ranking as survivors, computes their full four-term SQ16 distances, and chooses the final top-$K$ among them.

For $N$ probed candidates, a full byte-split scan computes four partial products per candidate, or $4N$ in total.
Rank-then-refine instead computes one partial product for every candidate and four only for the survivors, or $N+4RK$.
In addition, the INT8 ranking pass packs more coordinates into each tile than a BF16 pass.
Therefore, its bulk scan costs no more than a conventional low-precision scan over the same candidates.

\begin{figure}[t]
\centering
\includegraphics[width=\columnwidth]{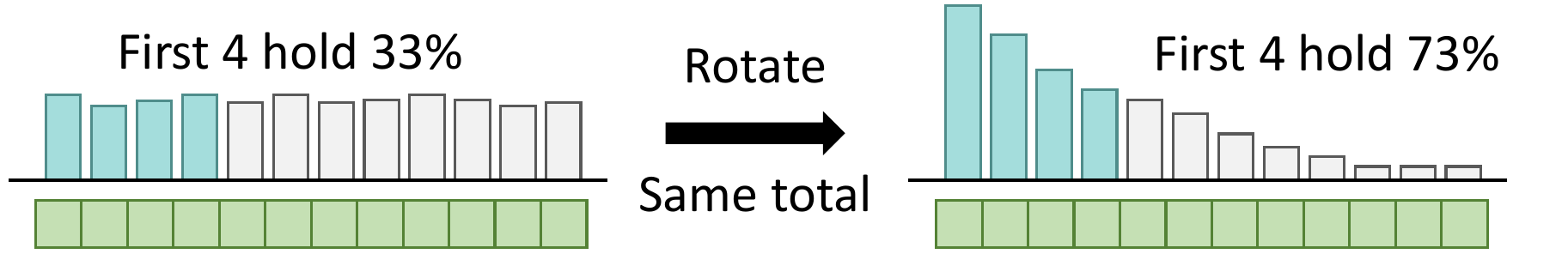}
\caption{An example of the rotation in the dimension cascade.\label{fig:dimc-rot}}
\end{figure}

\begin{figure}[t]
\centering
\includegraphics[width=\columnwidth]{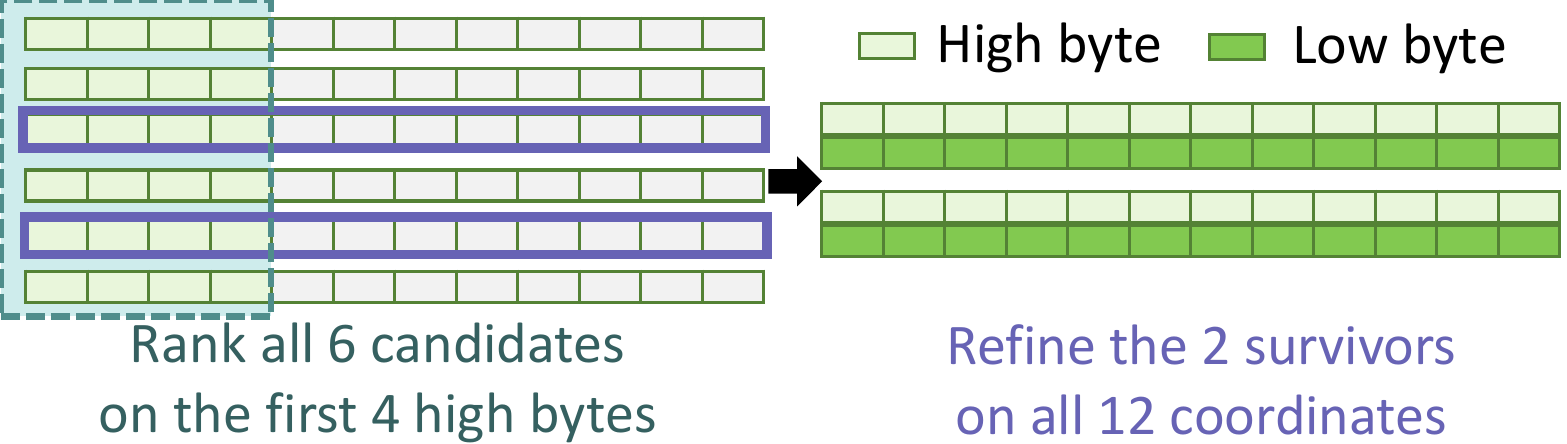}
\caption{An example of the dimension cascade.\label{fig:dimc-scan}}
\end{figure}

Figure~\ref{fig:cascade-ex} shows the two phases for one query.
With $N{=}32{,}000$ probed candidates, $K{=}10$, and $R{=}2$, Phase~1 computes one partial product for each candidate, and Phase~2 computes four for each of the $20$ survivors.
In total, the scan computes $32{,}080$ partial products instead of the $128{,}000$ of a full byte-split scan, about four times fewer.

\noindent\textbf{Dimension cascade.}
A naive prefix of the coordinates would discard useful signal arbitrarily.
At build time, before quantization, \sys{} applies an orthogonal rotation that concentrates variance into the leading coordinates and preserves full-dimensional distances.
At search time, \sys{} applies the same rotation to each query batch with one matrix multiplication, so the queries and the codes share one basis.
The ranking pass then uses only the first $d_0$ dimensions of the high-byte plane, which reduces the multiply-add operations for every candidate.
The refinement pass still uses all $d$ dimensions of the SQ16 code.
Therefore, the prefix affects only which candidates reach refinement, not the final distance.

Figures~\ref{fig:dimc-rot} and~\ref{fig:dimc-scan} show the dimension cascade with $d{=}12$ coordinates and a prefix of $d_0{=}4$.
As Figure~\ref{fig:dimc-rot} shows, the variance spreads evenly before the rotation, and the first four coordinates hold only about a third of it.
After the rotation, the same total variance concentrates in the leading coordinates, and the first four hold about three quarters of it.
In Figure~\ref{fig:dimc-scan}, the ranking pass reads only the four leading high bytes of all six candidates, a third of the multiply-add operations of the full ranking pass.
The refinement then reads all twelve coordinates, both the high and the low bytes, but only for the two survivors.

\noindent\textbf{Calibration.}
For each index and recall target, the calibration sweeps $\text{nprobe}$ per scan mode on a sample of queries with known true neighbors, with the full byte-split scan as the precision reference.
The probe count sets coverage by choosing which clusters are scanned, whereas the mode's survivor count sets how many candidates within those clusters receive the full SQ16 distance.
For high-dimensional data, the calibration also chooses the prefix length $d_0$.
A shorter prefix reduces ranking cost but may need more survivors, because its ranking is less accurate.
Ranking on all coordinates is the special case $d_0=d$, which the calibration selects when the leading rotated coordinates do not hold most of the variance.
The calibration selects the cheapest setting that reaches the target recall, and this setting may skip refinement entirely.

\subsection{Keeping the Matrix Engine Busy
\label{sec:design-busy}
\label{sec:design-feed}
\label{sec:design-coarse}}

Rank-then-refine reduces the multiplications per candidate, but the engine runs at full rate only if the rest of the pipeline keeps up.
A per-query IVF loop falls short in three ways.
First, it reads each list once per query.
Second, it stores the coordinates of each vector together, which does not match the layout of a tile.
Third, its centroid scan and top-$K$ updates run as scalar code between tiles, which a slow scan hides but a fast scan exposes.
\sys{} addresses the first with cluster-to-query inversion, as prior matrix-engine search does (\S\ref{sec:motiv-amx}).
It addresses the second with a VNNI-order code layout, and the third with a batched coarse search and top-$K$ selection kept off the critical path.

\begin{figure}[t]
\centering
\includegraphics[width=\columnwidth]{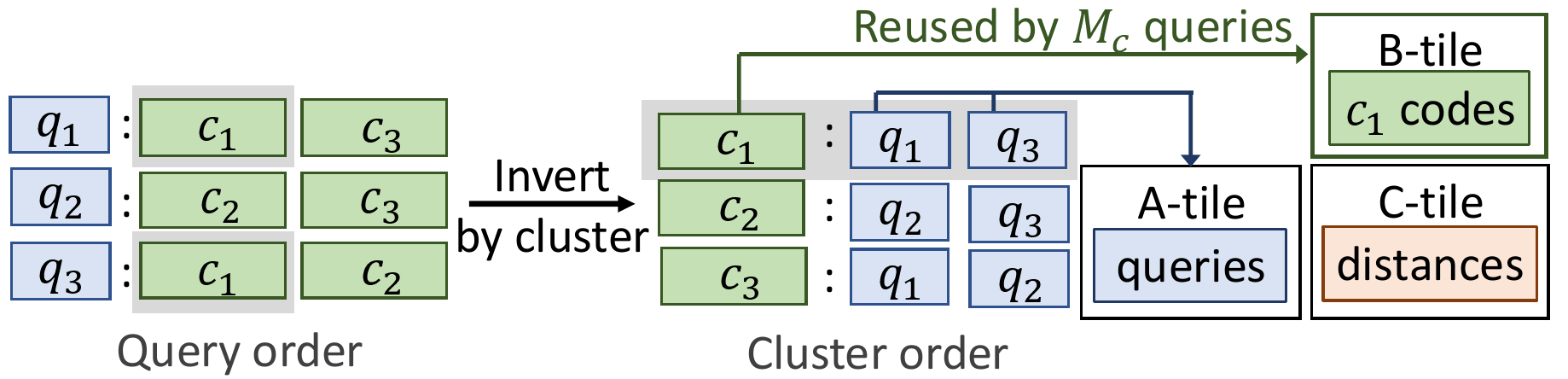}
\caption{An example of cluster-to-query inversion.\label{fig:inversion}}
\vspace{-3mm}
\end{figure}
\begin{figure}[t]
\centering
\includegraphics[width=\columnwidth]{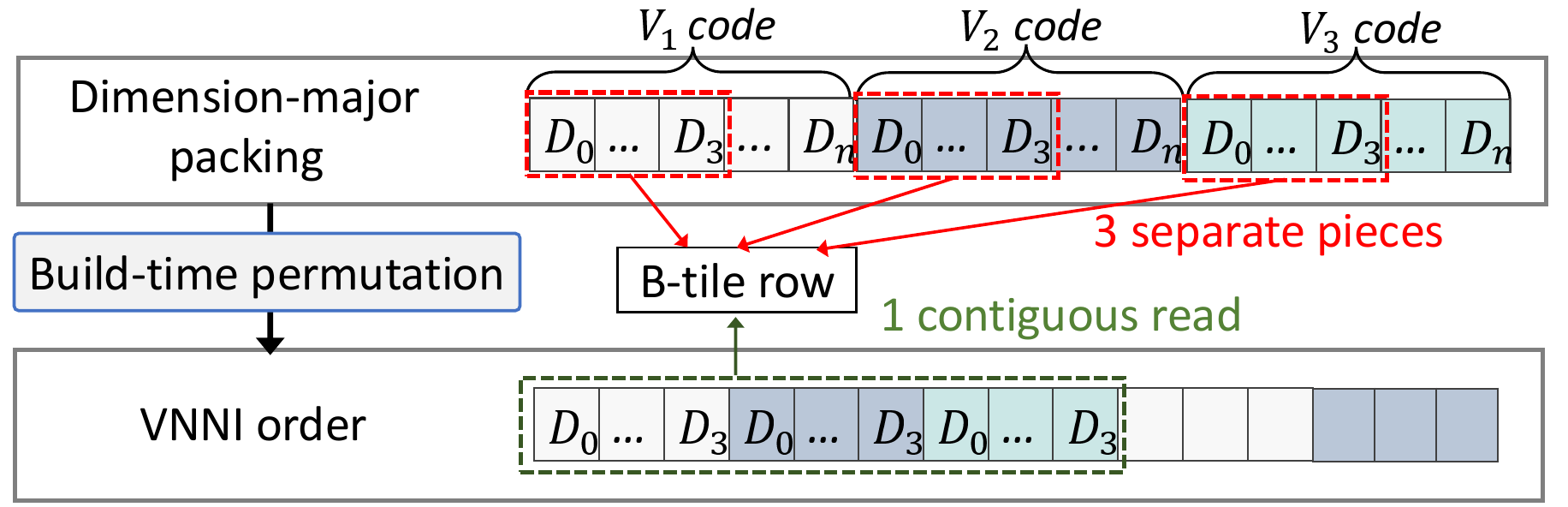}
\caption{An example of the VNNI-order code layout.\label{fig:vnni}}
\end{figure}

\noindent\textbf{Cluster-to-query inversion.}
\sys{} inverts each batch from query order to cluster order and scans each cluster $c$ once against the $M_c$ queries that probe it.
For a full batch of $B$ queries and uniform cluster popularity,
\begin{equation}
M_c \approx \frac{B\cdot \text{nprobe}}{\text{nlist}},
\label{eq:mc}
\end{equation}
where $\text{nlist}$ is the number of clusters and $M_c$ is the $M_{\mathrm{eff}}$ of Eq.~\ref{eq:crossover} for cluster $c$.
For example, a batch of $4{,}096$ queries that each probe $64$ of $2{,}048$ lists gives each list about $128$ queries.
For small $M_c$, \sys{} uses shorter tiles instead of padding every cluster to a fixed size.
\sys{} also does not wait for a full batch, and the engine processes the queued queries whenever it becomes idle.

Figure~\ref{fig:inversion} shows three queries that each probe two of three clusters.
In query order, $q_1$ and $q_3$ would each read the codes of $c_1$ from memory.
After the inversion, \sys{} loads the codes of $c_1$ once and multiplies them with one A-tile that holds both $q_1$ and $q_3$, so each loaded block serves two queries instead of one.

\noindent\textbf{VNNI-order code layout.}
A dimension-major layout stores the coordinates of one vector consecutively, which suits a single-query SIMD loop.
However, a B-tile holds a few consecutive coordinates of many database codes, and dimension-major codes force the runtime to gather these pieces from far-apart addresses for every tile.
For example, each 64-byte row of an INT8 B-tile holds four consecutive coordinates of 16 vectors, and with $1{,}024$-dimensional codes these 16 pieces lie $1{,}024$ bytes apart.
\sys{} instead stores each cluster in the Vector Neural Network Instructions (VNNI) layout, whose innermost block matches one row of a B-tile, at the cost of one permutation pass per cluster at build time.
Each tile row is then one contiguous 64-byte read, and the same layout serves both the ranking pass and the refinement.

Figure~\ref{fig:vnni} shows three vector codes.
In dimension-major order, the first four coordinates of $V_1$, $V_2$, and $V_3$ lie in three separate pieces, and one B-tile row must gather them.
After the build-time permutation, these coordinates are adjacent in VNNI order, and the B-tile row is one contiguous read.

\begin{figure}[t]
\centering
\includegraphics[width=\columnwidth]{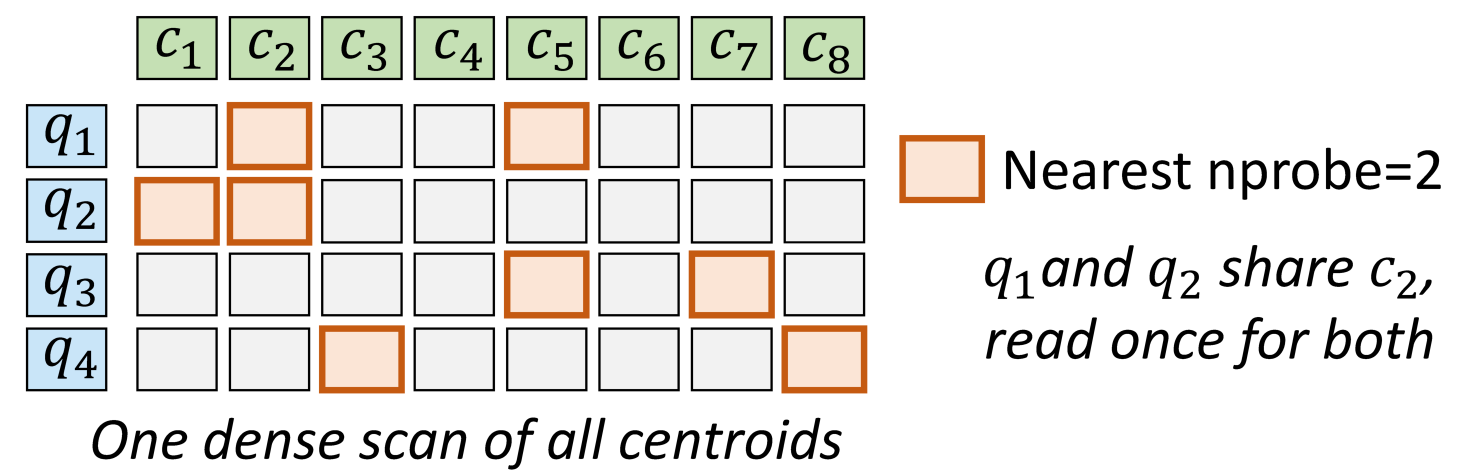}
\caption{An example of the batched coarse search.\label{fig:coarse-route}}
\vspace{-3mm}
\end{figure}

\begin{figure}[t]
\centering
\includegraphics[width=\columnwidth]{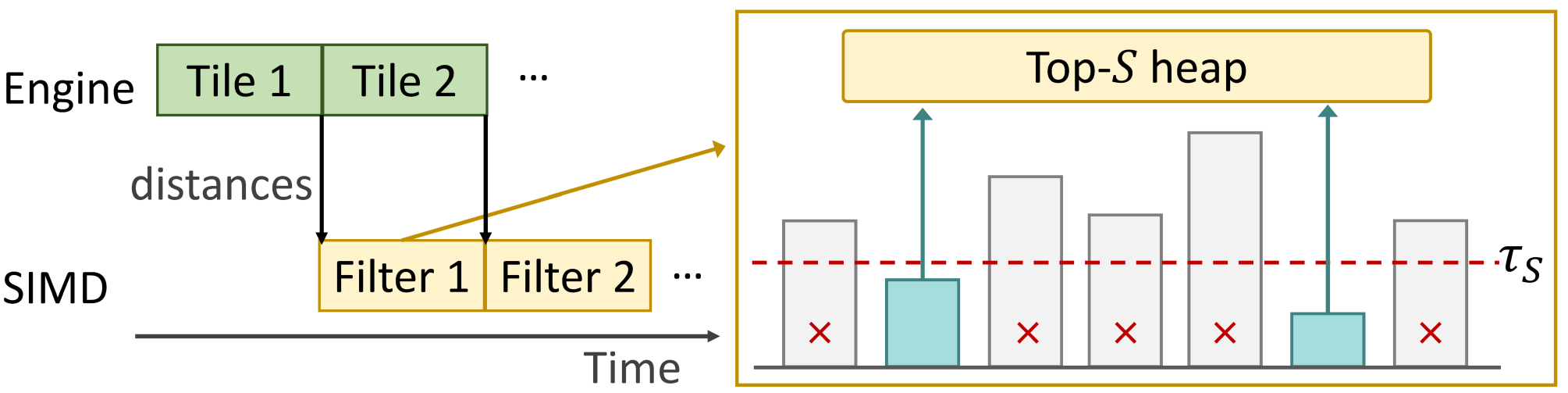}
\caption{An example of threshold-guarded selection.\label{fig:coarse-select}}
\vspace{-3mm}
\end{figure}

\noindent\textbf{Batched coarse search.}
A second, approximate index over the centroids can reduce the coarse-search cost, but it adds query-dependent control flow and can miss the nearest clusters.
\sys{} instead scores all centroids with a dense batched SQ16 scan, which is feasible because the centroid table is small.
For example, $2{,}048$ centroids of $1{,}024$ dimensions take 4\,MB as SQ16 codes, whereas one million such vectors take 2\,GB.
The centroids use the same byte planes and layout as the database codes.
Because every centroid is scored under the same SQ16 distance as the fine scan, this step adds no error from an approximate index.
Its cost also does not depend on $\text{nprobe}$, because every query scores all centroids.

Figure~\ref{fig:coarse-route} shows the coarse search for four queries and eight centroids.
One dense scan fills the whole distance matrix, and each query keeps its two nearest centroids.
Queries $q_1$ and $q_2$ both keep $c_2$, and the fine scan later reads the codes of $c_2$ once for both of them.

\noindent\textbf{Selection off the critical path.}
\sys{} uses two selection paths, depending on the probe count.
At high probe counts, the current $S$-th-best distance $\tau_S$ of each query drops quickly.
A SIMD compare then rejects distances larger than $\tau_S$ before they reach the heap, and it runs on vector units while the matrix engine produces the next tiles.
At low probe counts, $\tau_S$ stays high, and the compare rejects few distances.
In this case, \sys{} keeps a small insertion-sorted array of the best $S=R\!\cdot\!K$ candidates per cluster and merges these lists across clusters.
This local selection is exact, because the global top-$S$ of a query must lie in the union of its per-cluster top-$S$ sets.

In Figure~\ref{fig:coarse-select}, one vector compare discards the five of seven distances that lie above $\tau_S$.
Only the two distances below $\tau_S$ reach the heap, while the engine computes the next tile.

\subsection{Discussion
\label{sec:design-discussion}}

\noindent\textbf{Hardware portability.}
The \sys{} prototype targets Intel AMX, but the design is not specific to it.
It requires a low-precision matrix engine, a way to feed it dense tiles, and enough accumulator precision to recombine byte-split partial products safely.
Such engines are becoming a standard CPU feature.
For example, Armv9 CPUs provide a comparable unit through the Scalable Matrix Extension (SME).
Therefore, the byte split and rank-then-refine should apply to these engines, subject to their operand types and accumulator widths.

\noindent\textbf{Scaling beyond a single server.}
\sys{} is a per-shard, in-memory ANN search system, and larger datasets can be partitioned across multiple servers.
Two properties of the design carry over to this setting.
First, \sys{} reduces the cost of each distance comparison, and this advantage does not depend on the total dataset size.
Therefore, we expect the per-shard advantage to remain as more shards are added.
Second, a server stays compute-bound only with a large batch, and sharding the database keeps this batch, because every query that probes a list still reaches the server that holds the list.

\section{Implementation \label{implementation}}

\sys{} is implemented in about $9{,}000$ lines of C++17 and $2{,}000$ lines of kernel code written with x86 intrinsics.
A one-time calibration selects the scan mode and its parameters for each built index, and serving performs no per-query mode search.

\noindent\textbf{Index build.}
The builder trains the centroids with k-means on a sample of the vectors.
On high-dimensional data, it computes the rotation of the dimension cascade from the principal components of a sample.
It then assigns each vector to its nearest centroid and emits the SQ16 codes, the high-byte norms, and the VNNI-order lists in one streaming pass over the input.

\noindent\textbf{Kernels.}
The scan kernels use AMX tile instructions directly.
The full byte-split scan, the coarse search, and both forms of rank-then-refine share one kernel template and differ only in the byte planes and dimensions that they feed to the tile multiply.
The scan is cache-blocked.
It reads each inverted list in cache-sized blocks of codes and multiplies every 16-query tile against a block before it moves to the next block.
Therefore, each code is read from memory once while its block stays in cache.
On CPUs without AMX, the same pipeline runs with AVX-512 kernels.

\noindent\textbf{Runtime.}
For each batch, \sys{} rebuilds the cluster-to-query map in place in $O(B\cdot\text{nprobe})$ time with an indexed bucket array.
Worker threads take clusters round-robin and keep their heaps and scratch buffers private.
Therefore, the scan path needs no inter-thread synchronization.

\begin{table}[t]\centering\small
\caption{Datasets used in evaluation.
\label{tab:datasets}}
\setlength{\tabcolsep}{4pt}
\begin{tabular}{l l l r r}
\hline
\textbf{Name} & \textbf{\#Vectors} & \textbf{Type} & \textbf{Dim} & \textbf{\#Queries} \\
\hline
DEEP-96     & $10$M       & FP32  & $96$   & $10{,}000$ \\
Cohere-768  & $1$M /$10$M & FP32  & $768$  & $1{,}000$ \\
Cohere-1024 & $5$M        & FP32  & $1024$ & $10{,}000$ \\
BioASQ-1024 & $10$M       & FP32  & $1024$ & $3{,}106$ \\
OpenAI-1536 & $1$M / $5$M  & FP32  & $1536$ & $10{,}000$ / $1{,}000$ \\
OpenAI-3072 & $1$M        & FP32  & $3072$ & $10{,}000$ \\
\hline
Text2Image  & $100$M      & FP32  & $200$  & $10{,}000$ \\
\hline
BioASQ-QA   & $14.9$M     & FP32  & $1024$ & $5{,}729$ \\
HousingQA   & $1.8$M      & FP32  & $1024$ & $6{,}853$ \\
\hline
\end{tabular}
\vspace{-3mm}
\end{table}

\begin{figure*}[t]\centering
\includegraphics[width=\textwidth]{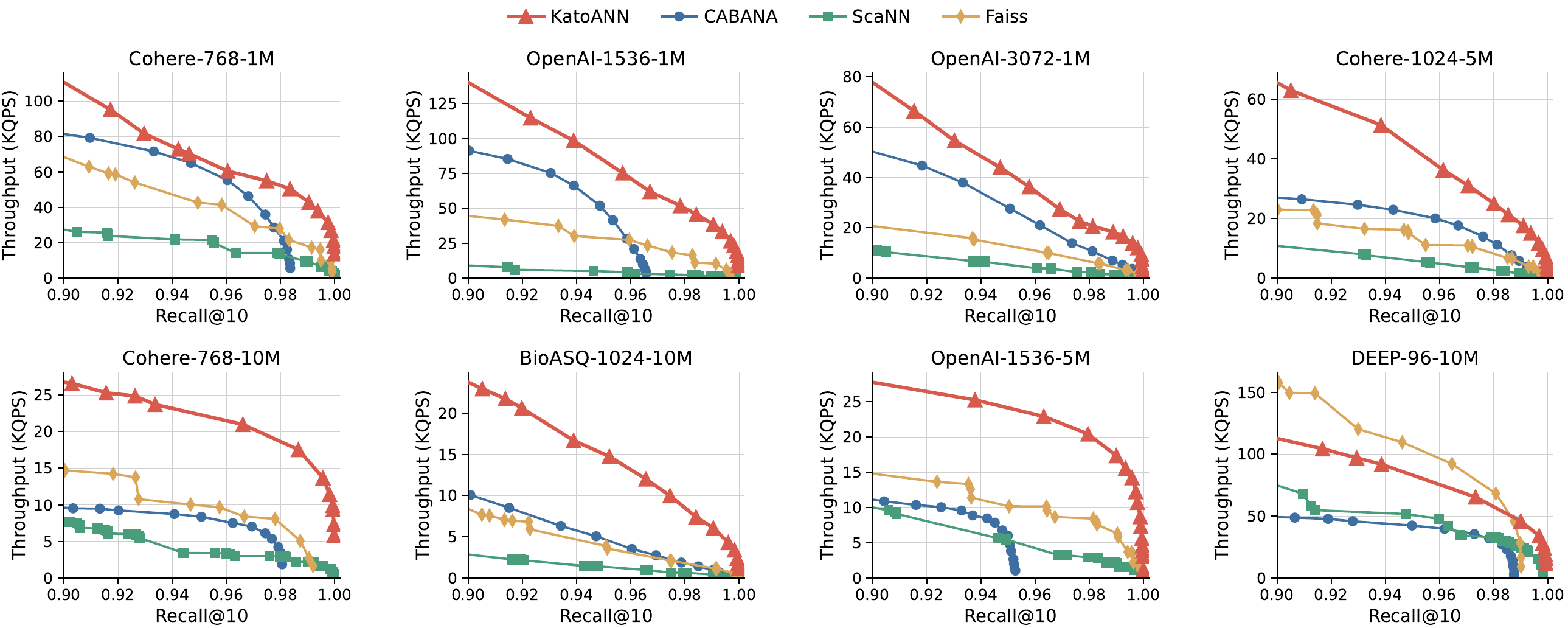}
\caption{Recall-throughput of partition-based systems on eight datasets.
\label{fig:eval-partition-pareto}}
\end{figure*}
\section{Evaluation \label{evaluation}}

\subsection{Methodology \label{sec:eval-method}}

\noindent\textbf{Testbed.}
We run the experiments on a CloudLab~\cite{cloudlab} c6620 node with one Intel Xeon Gold 5512U CPU ($28$ cores with AMX, simultaneous multithreading disabled), $128$\,GB of DDR5 memory, and NVMe storage.
The server runs Ubuntu $24.04$ with Linux kernel $6.8$.
Every system runs with $28$ threads.

\noindent\textbf{Datasets.}
Table~\ref{tab:datasets} summarizes eight datasets for the main experiments, one $100$M-vector dataset, and two for the RAG application.
DEEP-96~\cite{deep1b} is a low-dimensional benchmark, while the others contain $768$- to $3072$-dimensional text embeddings with up to $10$M vectors.
Cohere-768, OpenAI-1536-$5$M, and BioASQ-1024 are from VectorDBBench~\cite{vdbbench}, using embeddings of Wikipedia, C4, and BioASQ biomedical articles, respectively.
Cohere-1024 embeds Wikipedia using Cohere embed-multilingual-v3, while OpenAI-1536-$1$M and OpenAI-3072-$1$M embed DBpedia using text-embedding-ada-002 and text-embedding-3-large, respectively.
All text embeddings use cosine similarity, implemented as $L_2$ search over normalized vectors.
Text2Image-$100$M is the first $100$M vectors of the Yandex Text-to-Image benchmark~\cite{bigann21}, searched by inner product.
Ground truth is the exact FP32 top-$10$, and the queries come from each dataset's query set or a random holdout.
For the RAG application, we use BioASQ-QA~\cite{beir,bioasq} (PubMed articles) and HousingQA~\cite{housingqa} (U.S. state statutes), both embedded using BGE-M3~\cite{bgem3,pubmedmv}.
Because of this embedding model, BioASQ-QA is a different vector set from BioASQ-1024.

\noindent\textbf{Compared systems.}
We compare against partition-based and graph-based systems, as well as dimension-pruning and compact-code methods.
\begin{itemize}[leftmargin=*,nosep]
\item \textbf{Partition-based systems.}
The partition-based systems are CABANA, ScaNN, and Faiss FastScan.
CABANA~\cite{cabana} is the prior AMX-based ANN system.
ScaNN~\cite{scann} is the AVX-512 partition baseline.
Faiss~\cite{faiss} runs FastScan, which scans 4-bit PQ codes of an inverted file and re-ranks with FP32 vectors.

\item \textbf{Graph-based systems.}
The graph-based systems are Vamana~\cite{diskann} and HNSW~\cite{hnsw} over FP32 vectors, SVS~\cite{svs-lvq} with LVQ codes, which scalar-quantize each vector with its own range, and LeanVec~\cite{leanvec} codes, which apply LVQ to a lower-dimensional projection, and the disk-resident DiskANN~\cite{diskann} and PipeANN~\cite{pipeann}.

\item \textbf{Dimension pruning and compact codes.}
We additionally compare dimension pruning (ADSampling~\cite{adsampling}, DADE~\cite{dade}, and PDX~\cite{kuffo2025pdx}) and compact codes (extended RaBitQ~\cite{rabitq-ext}, CAQ~\cite{saq}, and SAQ~\cite{saq}).
All of them except PDX run on the inverted file of \sys{}, and PDX trains its own partition.
Their parameters are selected on a separate set of calibration queries.
\end{itemize}

\noindent\textbf{\sys{} configuration.}
At each recall target, \sys{} runs the fastest of its scan modes chosen by the calibration, namely the high-byte-only scan, rank-then-refine with $20$ survivors, the dimension cascade with $32$, and the full byte-split scan.
It takes queries in batches of $4{,}096$ or $10{,}000$ as CABANA does, and the systems that share its inverted file use $2{,}048$ lists at $1$M, $8{,}192$ at $5$M--$10$M, and $16{,}384$ at $100$M.

\noindent\textbf{Metrics.}
We report recall@$10$, throughput in queries per second (QPS), and serving memory as the steady-state resident set size (RSS) of the process, with the memory freed after loading returned to the OS.
We measure throughput by issuing the query set back to back in batches.

\begin{figure*}[t]\centering
\includegraphics[width=0.98\textwidth]{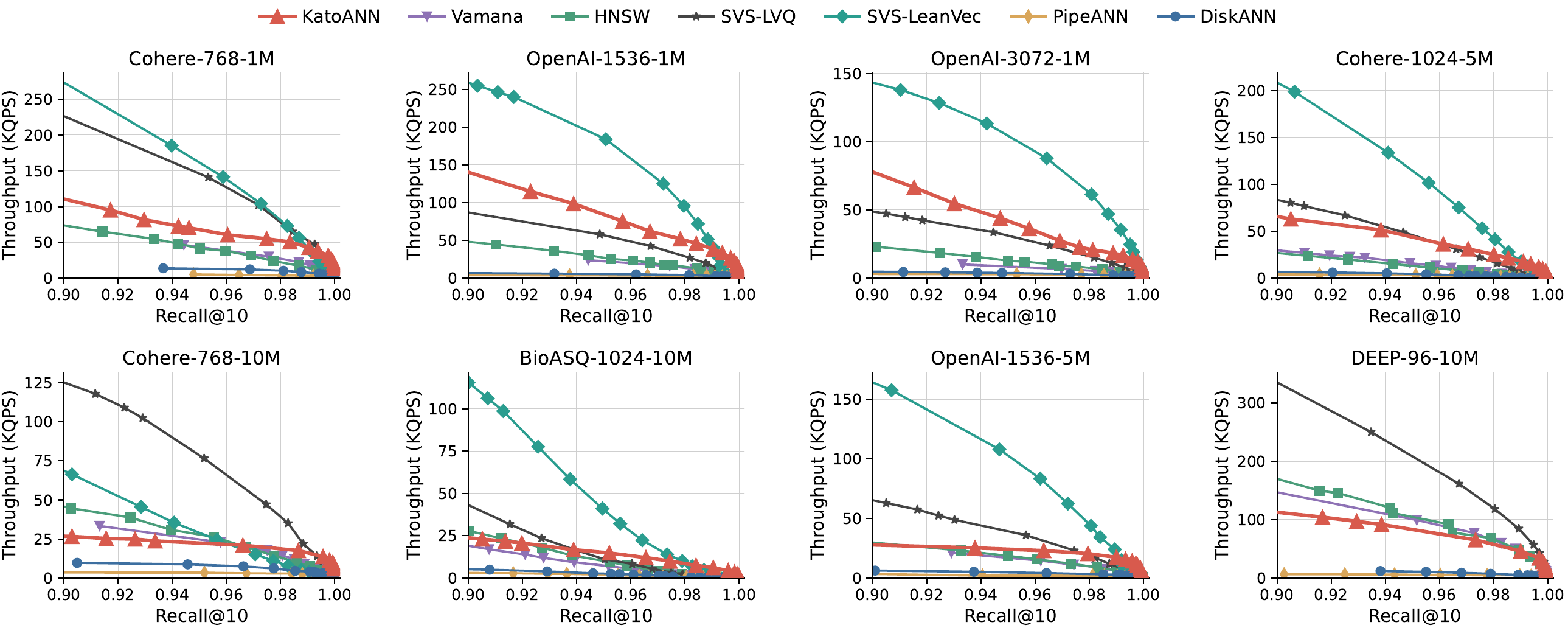}
\caption{Recall-throughput against graph-based systems on all eight datasets.
\label{fig:eval-pareto}}
\end{figure*}

\subsection{Main Results \label{sec:eval-main}}

\noindent\textbf{Comparison with partition-based systems.}
Figure~\ref{fig:eval-partition-pareto} shows that \sys{} has the highest throughput among partition-based systems across the high-recall range on all eight datasets.
On OpenAI-3072-$1$M at recall@$10\ge0.99$, \sys{} serves $16{,}624$\,QPS and has higher throughput than CABANA and ScaNN by $3.13\times$ and $11.7\times$, respectively.
Because CABANA uses the same inverted file, this gap comes from the scan, where \sys{} computes byte-split SQ16 distances and CABANA scans lossy BF16 codes.

The lossy baselines also stop short of high recall.
CABANA does not reach recall $0.99$ on five of the eight datasets, and Faiss FastScan does not reach $0.999$ on three, whereas \sys{} reaches $0.99$ on every dataset and extends to $0.999$.
Where ScaNN and FastScan reach $0.99$, \sys{} has higher throughput than them by $1.9\times$ to $31\times$ and $2.5\times$ to $5.4\times$, respectively.
This is because they reach high recall only by re-ranking against a resident full-precision copy in a second pass.
This result shows that \sys{} reaches high recall without giving up throughput.

\noindent\textbf{Crossover with graph-based systems.}
Figure~\ref{fig:eval-pareto} compares \sys{} with graph-based systems on all eight datasets.
Unlike the partition-based case, the result is a crossover rather than a uniform lead, because the matrix engine speeds up the IVF scan but not the graph walk.
At recall@$10\ge0.99$, \sys{} has higher throughput than SVS-LeanVec, the best reduced-dimension graph, by $1.10\times$ to $3.53\times$ on four of the seven high-dimensional datasets, and it leads on all seven at recall $0.999$.
Against the FP32 graphs, the margins grow to $1.8\times$ to $9.2\times$.
The largest exception at recall $0.99$ is OpenAI-3072-$1$M, the dataset whose variance is most concentrated in a few directions, where the small reduced-dimension codes of LeanVec make the graph cheaper and \sys{} leads only at recall $0.999$.

The crossover depends on two factors.
First, \sys{} computes each comparison faster, with a compute-bound matrix multiplication instead of a memory-bound gather.
Second, a graph compares the query against fewer vectors.
Higher dimensions make each comparison more expensive, which favors \sys{}.
On the low-dimensional DEEP-96 dataset, comparisons are cheap, and the graphs lead except at the extreme recall tail.
The disk-resident DiskANN and PipeANN trail the fastest in-memory graph throughout.
This result shows that \sys{} matches or outperforms graph-based systems on high-dimensional embeddings at high recall.

\begin{figure}[t]\centering
\includegraphics[width=0.95\linewidth]{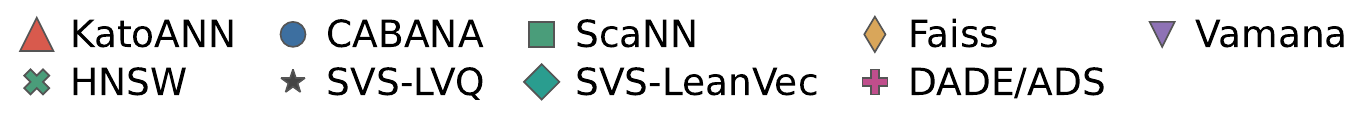}\\[-2pt]
\begin{minipage}[t]{0.49\linewidth}\centering
\includegraphics[width=\linewidth]{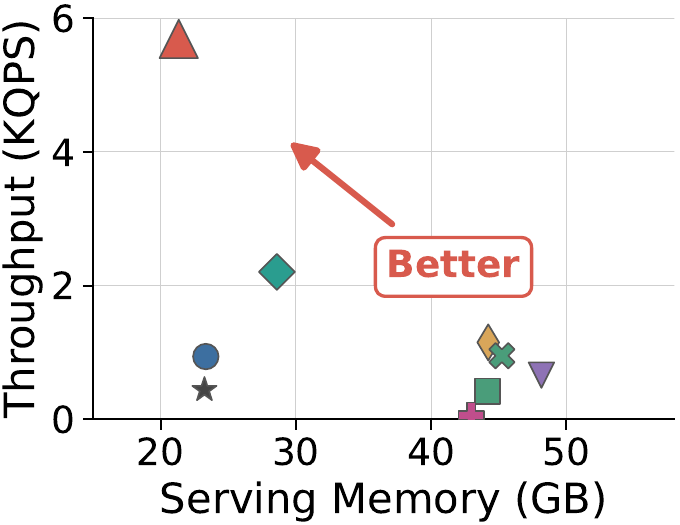}
\caption{Memory vs. Tput.
\label{fig:eval-mem-qps}}
\end{minipage}\hfill
\begin{minipage}[t]{0.49\linewidth}\centering
\includegraphics[width=\linewidth]{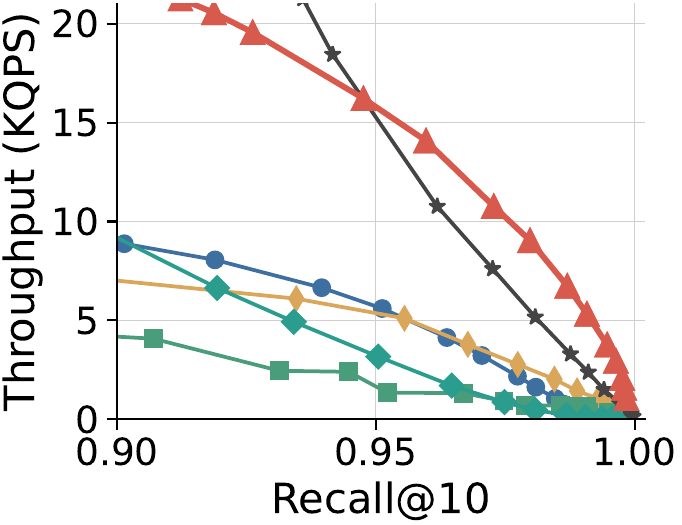}
\caption{Tput with $100$M.\label{fig:eval-t2i}}
\end{minipage}
\vspace{-3mm}
\end{figure}

\noindent\textbf{Memory footprint.}
Figure~\ref{fig:eval-mem-qps} plots serving memory against throughput at recall@$10\ge0.99$ on BioASQ-1024-$10$M.
\sys{} serves $5{,}691$\,QPS in $21.3$\,GB, the smallest footprint among the tested systems, and is the only system in the corner of low memory and high throughput.
The systems that keep FP32 vectors resident use $43$ to $48$\,GB, $2.0\times$ to $2.3\times$ as much, and the FP32 graphs among them run below $960$\,QPS.
The systems closest in footprint use more memory and are slower.
CABANA uses $23.3$\,GB but serves only $939$\,QPS, SVS-LVQ serves only $442$\,QPS in $23.2$\,GB, and SVS-LeanVec uses $28.6$\,GB and is slower than \sys{} by $2.6\times$.
This is because \sys{} keeps one $16$-bit code per vector and needs no full-precision copy for re-ranking.
This result shows that the single-copy design saves memory without giving up throughput.

\noindent\textbf{Performance on a 100M dataset.}
Figure~\ref{fig:eval-t2i} shows Text2Image-$100$M, which is ten times larger than the other datasets.
\sys{} and CABANA search its inner products as $L_2$ through an exact reduction that appends one coordinate to each vector, and builds that exceed memory use temporary swap, but every search runs without it.
\sys{} reaches recall@$10\ge0.99$ at $5{,}320$\,QPS and extends to $0.998$, whereas CABANA stops at $0.988$.
At recall@$10\ge0.99$, \sys{} has higher throughput than SVS-LVQ, Faiss, and ScaNN by $2.2\times$, $5.1\times$, and $8.4\times$, respectively, and SVS-LVQ is faster only below recall $0.95$.
SVS-LeanVec is slower by $47\times$ even with a projection learned from $100$K text queries~\cite{leanvec}, and HNSW does not fit in memory.
On this $200$-dimensional dataset, the calibration ranks on all coordinates ($d_0{=}d$).

\begin{figure}[t]
\centering
\includegraphics[width=0.7\linewidth]{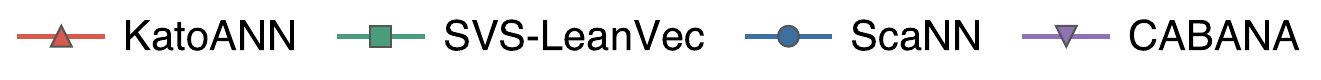}\\[-2pt]
\subfloat[Throughput]{%
\includegraphics[width=0.49\linewidth]{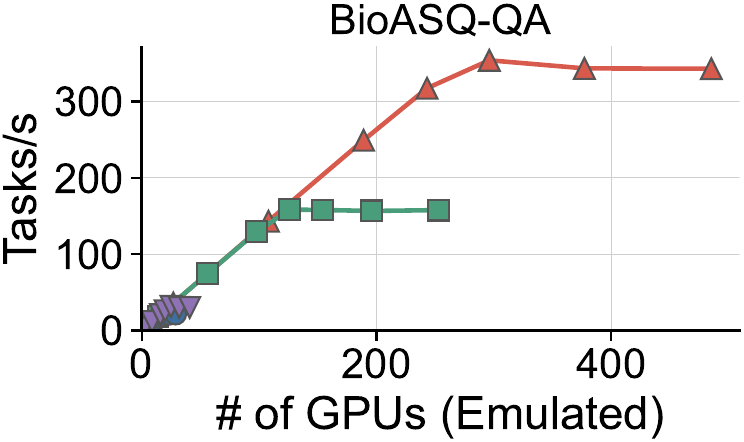}\hspace{0.02\linewidth}%
\includegraphics[width=0.49\linewidth]{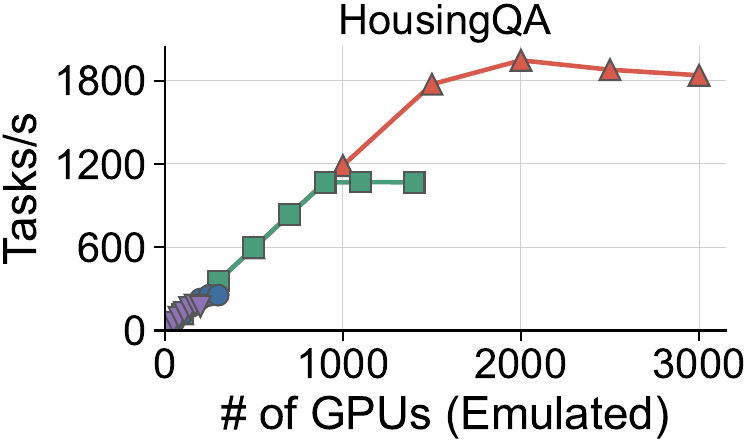}}\\
\subfloat[p99 latency]{%
\includegraphics[width=0.49\linewidth]{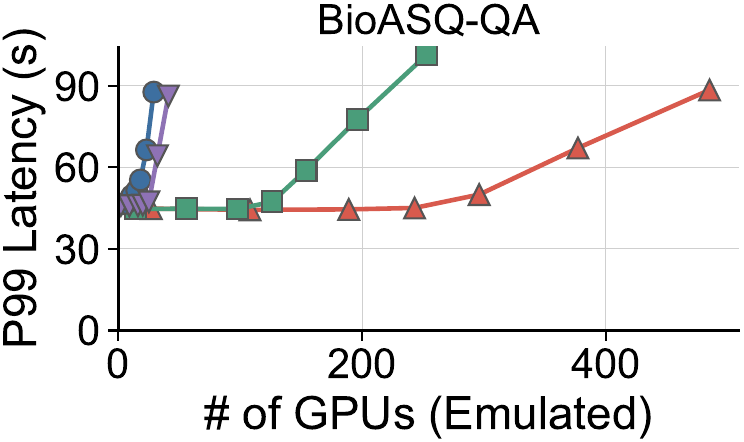}\hspace{0.02\linewidth}%
\includegraphics[width=0.49\linewidth]{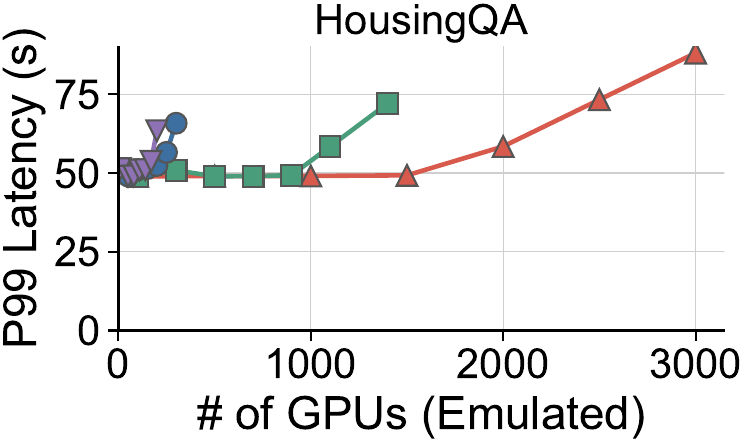}}
\caption{Agentic RAG with emulated GPUs at recall@$10\geq0.995$ (Qwen3-8B, one A30 per $32$ agents).}
\label{fig:eval-rag}
\end{figure}

\subsection{Application: Agentic RAG \label{sec:eval-rag}}
Agentic RAG needs high throughput, because many agents search at once, and high recall, because a missed passage can remove evidence that later reasoning steps need.
Therefore, we measure how many GPUs one CPU retrieval server can support at recall@$10\geq0.995$ on BioASQ-QA and HousingQA, the two RAG datasets in Table~\ref{tab:datasets}.

\noindent\textbf{Setup.}
Its LLM agents run on a CloudLab d7525 node with one NVIDIA A30 GPU, and its retrieval server runs on a c6620 node of the same configuration as the testbed.
Each A30 runs Qwen3-8B~\cite{qwen3} on vLLM~\cite{vllm} with the BGE-M3 query encoder, and each agent issues up to three search rounds of up to four queries per task.
The retrieval server runs each system at its highest-throughput setting that reaches the target recall.
One A30 reaches $99\%$ of its peak task throughput at $32$ concurrent agents.
To emulate more GPUs, we replay many copies of the traces of these agents, with their recorded LLM times, against the retrieval server and count every $32$ replayed agents as one GPU.

\noindent\textbf{Capacity of one retrieval server.}
Figure~\ref{fig:eval-rag} shows that task throughput grows linearly with the number of emulated GPUs until the retrieval server saturates.
After that point, queries wait at the server, and the 99th-percentile (p99) task latency rises steeply for every system.
We define the capacity of a server as the largest number of GPUs at which p99 task latency stays within $10\%$ of its single-GPU value, which is $44.4$\,s on BioASQ-QA and $49.0$\,s on HousingQA.
On BioASQ-QA, one \sys{} server supports $243$ GPUs, compared with $126$ for SVS-LeanVec, $25$ for CABANA, and fewer than $11$ for ScaNN.
On the smaller HousingQA dataset, every system supports more GPUs per server: $1{,}500$ GPUs for \sys{}, $900$ for SVS-LeanVec, $200$ for ScaNN, and $170$ for CABANA.

\begin{figure}[t]\centering
\includegraphics[width=0.84\linewidth]{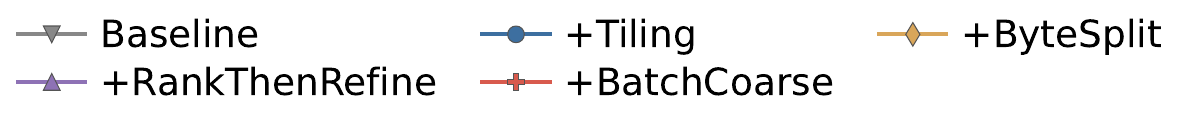}\\[-5pt]
\subfloat[BioASQ-1024-10M]{\includegraphics[width=0.50\linewidth]{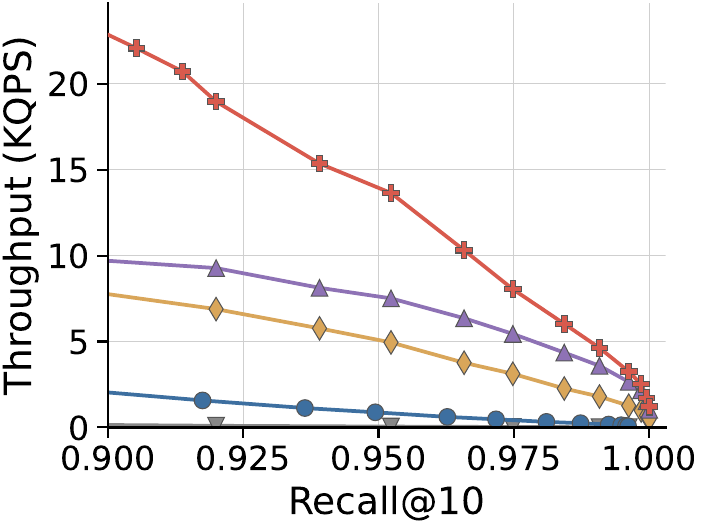}}\hfill
\subfloat[OpenAI-3072-1M]{\includegraphics[width=0.50\linewidth]{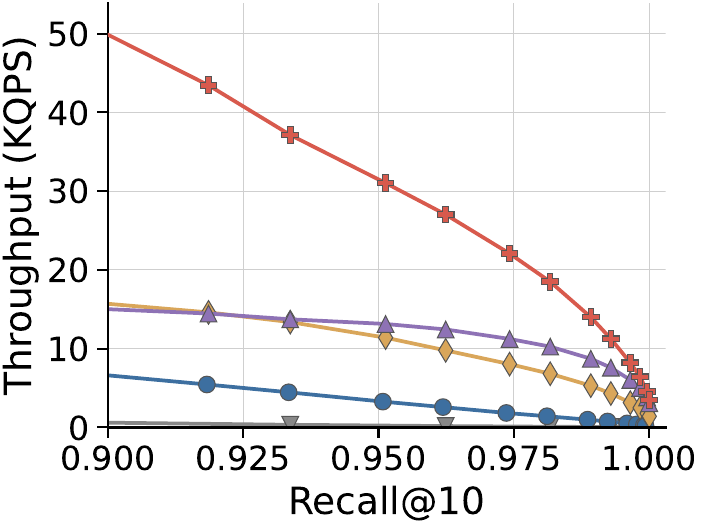}}
\caption{Cumulative ablation.}
\label{fig:eval-ablation}
\end{figure}

\noindent\textbf{Number of retrieval servers.}
A deployment scales its retrieval tier by replicating servers.
Because one \sys{} server supports more GPUs, a deployment needs proportionally fewer \sys{} servers for the same number of GPUs.
On HousingQA, it needs fewer servers than SVS-LeanVec by $1.7\times$ and fewer than ScaNN and CABANA by $7.5\times$ to $8.8\times$.
On BioASQ-QA, which requires a wider search, the gaps widen to $1.9\times$ and an order of magnitude, respectively.
Because \sys{} also stores each vector only once, these servers need no memory for a second full-precision copy.
This result shows that \sys{} shrinks the retrieval tier that a given number of GPUs needs at high recall.

\subsection{Deep Dive \label{sec:eval-deep}}

\noindent\textbf{Contribution of each technique.}
Figure~\ref{fig:eval-ablation} adds the techniques of \sys{} one at a time on BioASQ-1024-$10$M and OpenAI-3072-$1$M.
Cluster-batched tiling, which \sys{} shares with CABANA~\cite{cabana}, is the largest single step on BioASQ-1024-$10$M, about $14\times$ over the unbatched per-query scan at recall $0.99$, because the matrix engine runs at full rate only when its tiles are full.
On top of tiling, byte-split precision adds $10.7\times$ and $5.7\times$ on the two datasets, the rank-then-refine scan adds $2.00\times$ and $1.76\times$, and the batched coarse search adds $1.29\times$ and $1.48\times$.
In contrast, replacing AVX-512 with AMX without these techniques brings only about $2\times$.
At the final operating point on BioASQ-1024-$10$M, the AMX scan kernels and the overlapped threshold filtering take $86\%$ of the search cycles.
This result shows that the speedup comes from a pipeline that keeps the matrix engine busy, not from a fast distance kernel alone.

\begin{figure}[t]\centering
\includegraphics[width=0.93\linewidth]{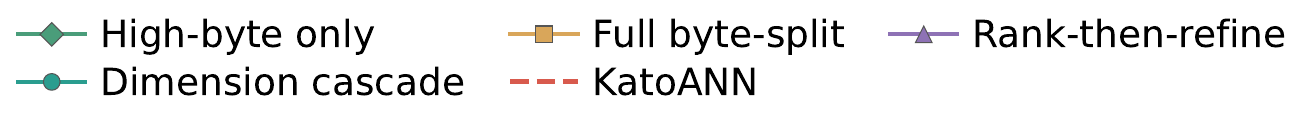}\\[-5pt]
\subfloat[BioASQ-1024-10M]{\includegraphics[width=0.50\linewidth]{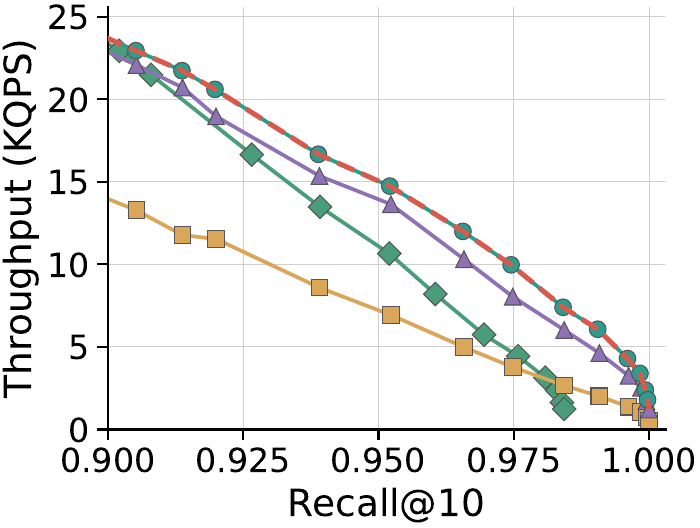}}\hfill
\subfloat[OpenAI-3072-1M]{\includegraphics[width=0.50\linewidth]{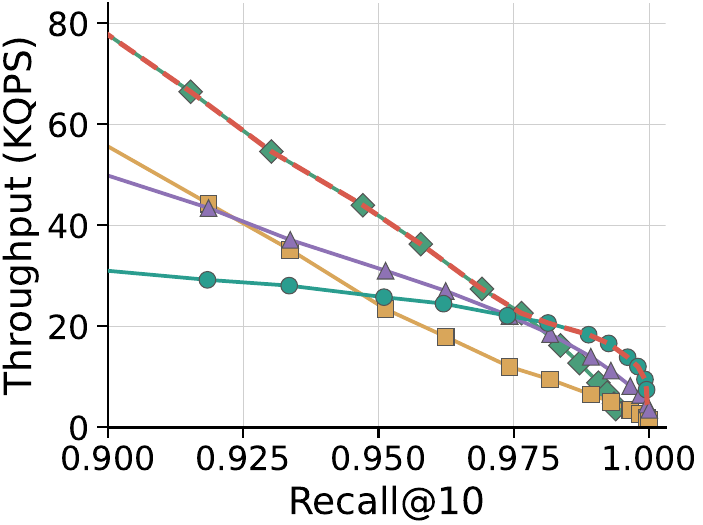}}
\caption{Impact of scan modes.}
\label{fig:eval-modes}
\end{figure}

\begin{figure}[t]\centering
\includegraphics[width=0.9\linewidth]{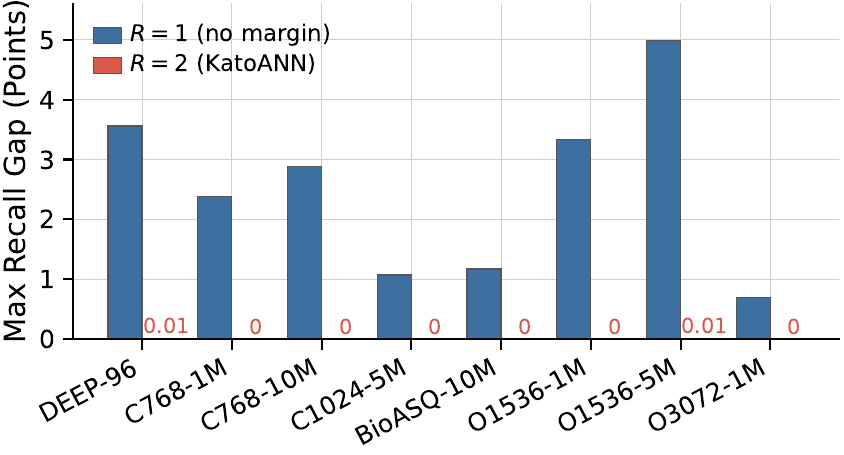}
\caption{Largest recall gap between rank-then-refine on all dimensions and the full byte-split scan, over $K\in\{10,100\}$ and $\text{nprobe}\in\{64,256,512\}$.
\label{fig:cascade-identity}}
\end{figure}

\noindent\textbf{Impact of the scan mode.}
Figure~\ref{fig:eval-modes} compares the scan modes of \sys{} on the same two datasets.
The high-byte-only scan is the cheapest, with one INT8 product per candidate, and the fastest at moderate recall, but its $8$-bit truncation limits recall to about $0.98$ and $0.994$, respectively.
The full byte-split scan reaches recall $0.999$, but it computes all four products for every candidate and is the slowest at matched recall.
Rank-then-refine reaches the same recall faster, because it computes the full distance only for the $R{\cdot}K$ survivors.
The dimension cascade, which ranks on a rotated prefix, is the fastest mode at high recall and serves $6{,}073$ and $16{,}624$\,QPS at recall $0.99$.
On OpenAI-3072-$1$M, it is $1.48\times$ faster than rank-then-refine on all dimensions, because its ranking pass reads only a short prefix of each $3{,}072$-dimensional vector.
No single mode is the fastest at every recall target, so \sys{} calibrates once per index and serves the fastest mode at each target.

\noindent\textbf{Recall of rank-then-refine.}
Rank-then-refine is a safe speedup only if its survivors keep the true neighbors, and the survivor factor $R$ sets how many there are, $R\!\cdot\!K$ per query (\S\ref{sec:design-cascade}).
Therefore, Figure~\ref{fig:cascade-identity} reports the largest recall gap between rank-then-refine and the full byte-split scan on all eight datasets, over $K\in\{10,100\}$ and $\text{nprobe}\in\{64,256,512\}$.
With $R{=}2$, that is, twice as many survivors as returned neighbors, rank-then-refine matches the recall of the full scan on every dataset, with a largest gap of $0.01$ percentage points.
With only $K$ survivors ($R{=}1$), recall drops by up to $5.0$ percentage points, because a true neighbor sometimes falls just below the top-$K$ of the high-byte ranking.
Larger factors ($R{=}3$ and $4$) return the same recall at lower throughput.
This result shows that rank-then-refine on all dimensions loses no recall and that $R{=}2$ is a saturation point rather than a sensitive threshold.

\begin{figure}[t]\centering
\centering
\includegraphics[width=0.76\linewidth]{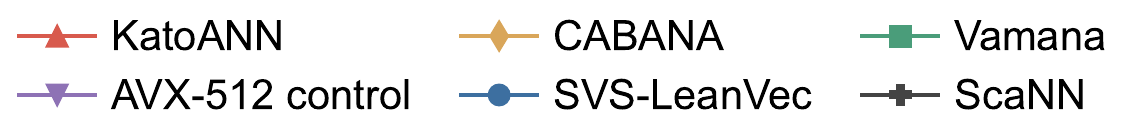}\\[-5pt]
\subfloat[OpenAI-3072-1M]{\includegraphics[width=0.50\linewidth]{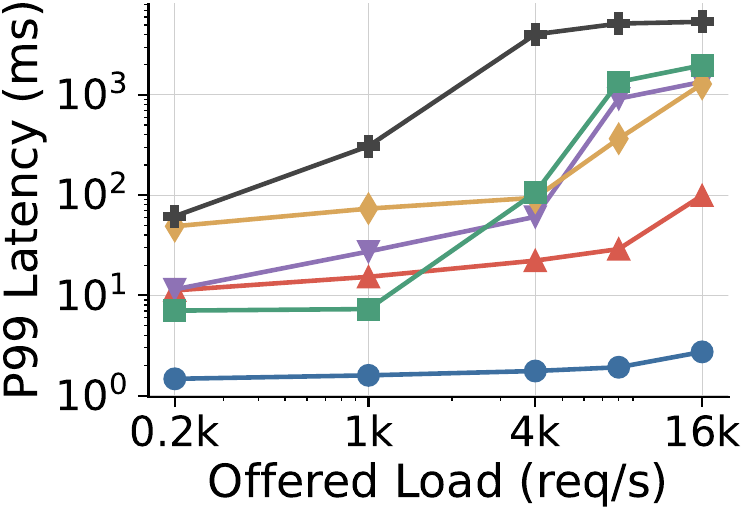}}
\subfloat[BioASQ-1024-10M]{\includegraphics[width=0.50\linewidth]{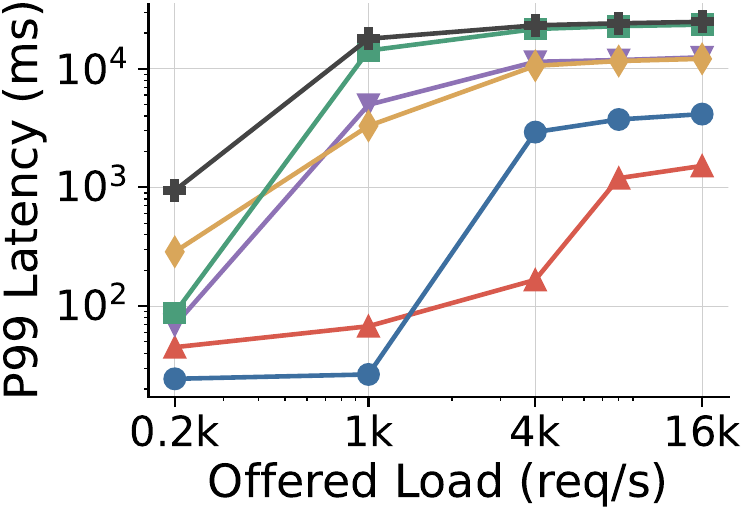}}
\caption{p99 search latency under load.}
\label{fig:eval-latency}
\end{figure}

\begin{figure}[t]\centering
\includegraphics[width=0.78\linewidth]{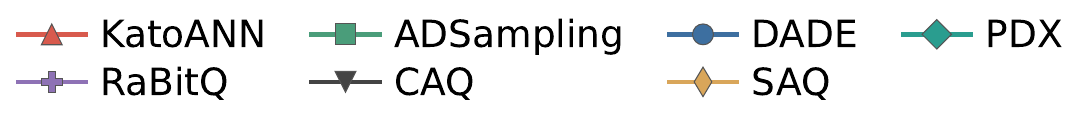}\\[-5pt]
\subfloat[OpenAI-3072-1M]{\includegraphics[width=0.50\linewidth]{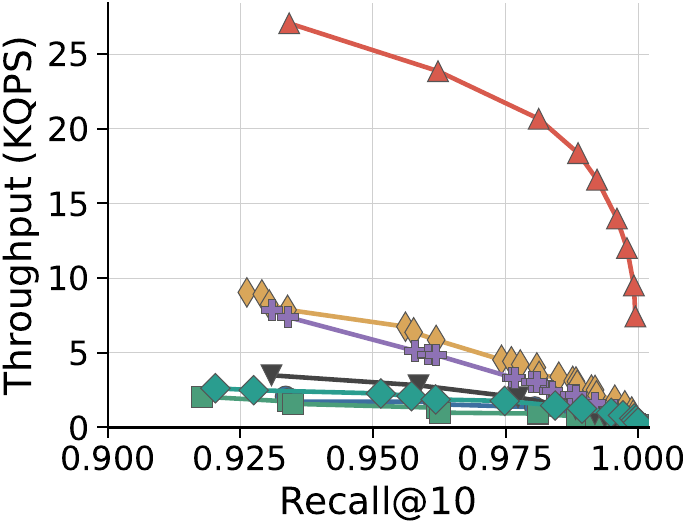}}\hfill
\subfloat[BioASQ-1024-10M]{\includegraphics[width=0.50\linewidth]{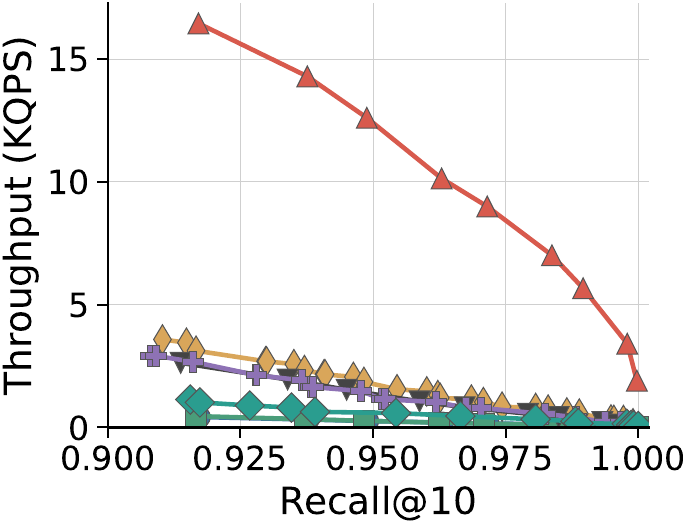}}
\caption{Recall-throughput vs.\ dimension pruning and compact codes.
\label{fig:eval-pruning}}
\end{figure}

\noindent\textbf{Tail latency under load.}
Figure~\ref{fig:eval-latency} shows that, among partition-based systems, \sys{} sustains the highest load at the lowest tail latency.
We replay a Poisson trace of $10{,}000$ requests at $200$ to $16{,}000$ requests per second, with each system at its lightest setting that reaches recall@$10\ge0.99$.
Batched systems take all waiting requests up to $B{=}4{,}096$, Vamana serves each request on its own worker, and latency includes queueing.
On BioASQ-1024-$10$M, \sys{} keeps p99 latency at $166$\,ms at $4{,}000$ requests per second, whereas CABANA, ScaNN, SVS-LeanVec, and Vamana saturate below $2{,}200$ requests per second and exceed $2.9$\,s.
The same pipeline on AVX-512 instead of AMX saturates at about $770$ requests per second, which shows that the matrix engine provides this capacity.
On OpenAI-3072-$1$M, where SVS is faster, SVS keeps p99 latency below $3$\,ms up to $4{,}000$ requests per second, while \sys{} reaches $22$\,ms.
This latency comes from batching, and it is small compared with the $44$ to $49$\,s of an agentic RAG task (\S\ref{sec:eval-rag}).

\noindent\textbf{Comparison with dimension pruning and compact codes.}
Figure~\ref{fig:eval-pruning} compares \sys{} with dimension pruning (ADSampling, DADE, and PDX) and compact codes (extended RaBitQ, CAQ, and SAQ), each in its best configuration.
These methods also reduce dimensions or bits, but per query rather than as batched matrix multiplications.
ADSampling and DADE compute only $1.8$ to $4.4\%$ of the dimensions, yet \sys{} is faster than them by $16\times$ to $28\times$ at recall $0.99$ on OpenAI-3072-$1$M and by $54\times$ to $69\times$ on BioASQ-1024-$10$M.
Against SAQ, the fastest of these methods, the gaps are $6.6\times$ and $10.6\times$.
This is because pruning runs a hypothesis test per candidate at every checkpoint, which keeps a scalar control loop on the critical path.
Compact codes read fewer bytes, but each query still scans its lists alone on SIMD units, and the scan stays bandwidth-bound.
\sys{} runs no per-candidate test and reuses each loaded code block across all the queries that probe it.
This result shows that reducing the dimensions or bits that each query reads does not replace a batched matrix scan.

\begin{figure}[t]\centering
\includegraphics[width=0.72\linewidth]{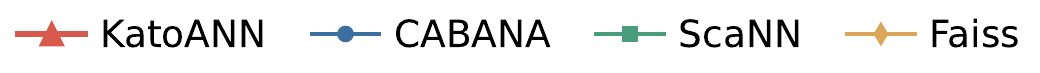}\\[-5pt]
\subfloat[OpenAI-3072-1M]{\includegraphics[width=0.50\linewidth]{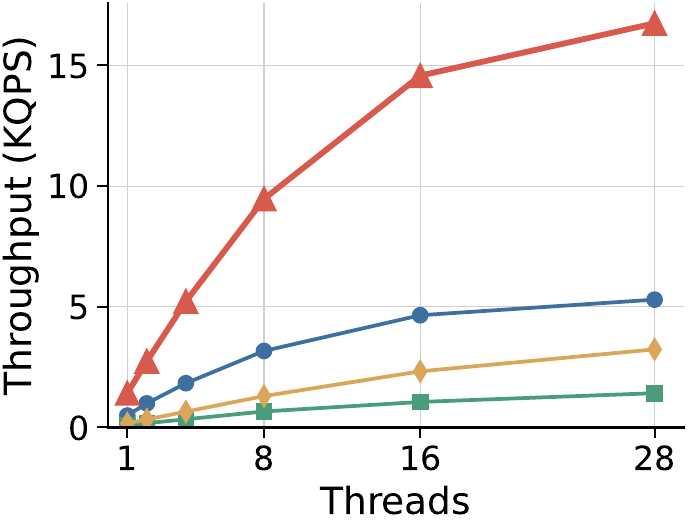}}\hfill
\subfloat[BioASQ-1024-10M]{\includegraphics[width=0.50\linewidth]{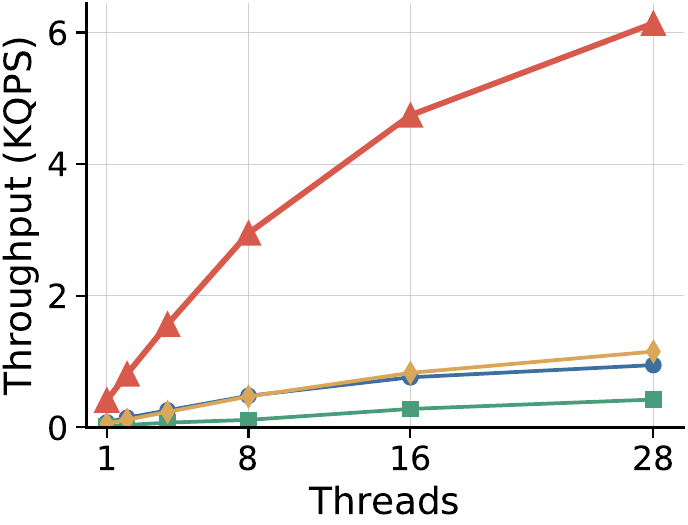}}
\caption{Thread scaling.
\label{fig:eval-thread}}
\end{figure}

\noindent\textbf{Thread scaling.}
Figure~\ref{fig:eval-thread} varies the number of threads from $1$ to $28$ at the recall-$0.99$ setting of each system.
On OpenAI-3072-$1$M, \sys{} scales by about $12\times$, because its heaps are private to each thread and its scan stages keep their data local.
CABANA, on the same inverted file, scales by $10.8\times$.
As a result, the lead of \sys{} over CABANA widens from $2.9\times$ at one thread to $3.2\times$ at $28$ threads.
At $28$ threads, \sys{} has higher throughput than ScaNN and Faiss FastScan by $11.8\times$ and $5.2\times$, respectively.
On BioASQ-1024-$10$M, \sys{} scales by $15\times$.
Because each system runs one fixed setting here, these ratios differ slightly from those on the Pareto curves.
This result shows that the lead of \sys{} holds as more cores serve queries.

\section{Related Work \label{relatedwork}}

\noindent\textbf{Matrix-engine and accelerated search.}
ScaNN~\cite{scann}, Faiss FastScan~\cite{faiss,fastscan}, and AlayaLaser~\cite{alayaLaser} accelerate distance computations using SIMD, but process queries independently without exploiting cross-query reuse.
CABANA~\cite{cabana} batches IVF cluster scans onto AMX but scans lossy BF16 codes, whereas \sys{} ranks and refines from one SQ16 code.
RAIRS~\cite{rairs} also groups queries by IVF list for cache reuse and stores full blocks shared by redundant list assignments once.
GPU-based ANN search systems~\cite{cagra,bang,gustann,flashanns} move the scan to a separate, expensive device.
\sys{} instead targets the matrix engine already present in the CPU.

\noindent\textbf{Quantization, refinement, and distance pruning.}
PQ~\cite{pq}, anisotropic quantization~\cite{scann}, LVQ and LeanVec~\cite{svs-lvq,leanvec}, and RaBitQ~\cite{rabitq,rabitq-ext} reduce storage and distance costs, and SVS-LVQ and SymphonyQG~\cite{symphonyqg} pair quantization with graphs.
SAQ~\cite{saq}, extended RaBitQ~\cite{rabitq-ext}, and LeanVec-Sphering~\cite{gleanvec} rank on part of a code and refine with the full code.
ADSampling~\cite{adsampling}, DADE~\cite{dade}, and TRIM~\cite{trim} stop distance computations early, and DARTH~\cite{darth} and Ada-ef~\cite{adaef} adapt how long each query searches.
These approaches remain query-centric, whereas \sys{} draws both distances from one SQ16 code and reuses each code across a batch of queries.

\noindent\textbf{Graph, storage, and database systems.}
Graph indexes~\cite{hnsw,diskann,pipeann} reduce candidate comparisons, complementing \sys{}'s cheaper distance computation.
HARMONY~\cite{harmony} and CoTra~\cite{cotra} distribute an index across machines and coordinate search over the network.
Storage-oriented search~\cite{spann,starling,margo,pageann,smartanns,fusionanns}, filtered search~\cite{acorn,curator,sieve-ann,navix,filtered-diskann,ung,nhq,vecflow,postgres-fvs,nemoann}, and adaptive partitioning~\cite{quake} address data access, predicates, and index maintenance.
Vector databases~\cite{milvus,pgvector,vbase} and RAG frameworks~\cite{langchain,flashrag} provide broader integration opportunities.
\sys{} can serve as the high-throughput distance engine inside such systems.

\section{Conclusion \label{conclusion}}
We presented \sys{}, a high-performance and high-recall ANN search system that leverages the capability of built-in CPU matrix engines.
To address recall degradation caused by the matrix engines, \sys{} stores each vector as a 16-bit code split into two bytes and builds near-exact distances from 8-bit matrix multiplications.
To minimize extra multiplication overhead, it computes the full distance only for the few survivors of a high-byte ranking.
Our evaluation shows that \sys{} is the fastest partition-based system and is comparable to graph-based systems, outperforming them at the highest recall targets.
We hope this work demonstrates that the hardware constraints of CPU matrix engines can be overcome through careful systems design.
\section*{Acknowledgement}
This research was sponsored by the National Research Foundation of Korea (NRF) grants funded by the Ministry of Science and ICT (No. RS-2025-00522990). Gyuyeong Kim is the corresponding author.

\bibliographystyle{plain}
\bibliography{./gykim,./katoann}

\begin{thebibliography}{10}

\bibitem{svs-lvq}
Cecilia Aguerrebere, Ishwar~Singh Bhati, Mark Hildebrand, Mariano Tepper, and
  Theodore Willke.
\newblock Similarity search in the blink of an eye with compressed indices.
\newblock {\em Proceedings of the VLDB Endowment (PVLDB)}, 16(11):3433--3446,
  2023.

\bibitem{fastscan}
Fabien Andr{\'e}, Anne-Marie Kermarrec, and Nicolas Le~Scouarnec.
\newblock Cache locality is not enough: High-performance nearest neighbor
  search with product quantization fast scan.
\newblock {\em Proceedings of the VLDB Endowment (PVLDB)}, 9(4):288--299, 2015.

\bibitem{arm-sme}
{Arm Ltd.}
\newblock {A}rm architecture reference manual supplement, the scalable matrix
  extension ({SME}), for {A}rmv9-a.
\newblock Document DDI 0616,
  \url{https://developer.arm.com/documentation/ddi0616/latest/}, 2023.

\bibitem{deep1b}
Artem Babenko and Victor Lempitsky.
\newblock Efficient indexing of billion-scale datasets of deep descriptors.
\newblock In {\em Proceedings of the 2016 IEEE Conference on Computer Vision
  and Pattern Recognition (CVPR)}, pages 2055--2063, 2016.

\bibitem{ung}
Yuzheng Cai, Jiayang Shi, Yizhuo Chen, and Weiguo Zheng.
\newblock Navigating labels and vectors: A unified approach to filtered
  approximate nearest neighbor search.
\newblock {\em Proceedings of the ACM on Management of Data (SIGMOD)},
  2(6):1--27, 2024.

\bibitem{langchain}
Harrison Chase.
\newblock {LangChain}.
\newblock \url{https://github.com/langchain-ai/langchain}, 2022.

\bibitem{darth}
Manos Chatzakis, Yannis Papakonstantinou, and Themis Palpanas.
\newblock {DARTH: Declarative Recall Through Early Termination for Approximate
  Nearest Neighbor Search}.
\newblock {\em Proceedings of the ACM on Management of Data}, 3(4), 2025.

\bibitem{bgem3}
Jianlv Chen, Shitao Xiao, Peitian Zhang, Kun Luo, Defu Lian, and Zheng Liu.
\newblock {M3-Embedding}: Multi-linguality, multi-functionality,
  multi-granularity text embeddings through self-knowledge distillation.
\newblock In {\em Findings of the Association for Computational Linguistics:
  ACL 2024}, pages 2318--2335, 2024.

\bibitem{spann}
Qi~Chen, Bing Zhao, Haidong Wang, Mingqin Li, Chuanjie Liu, Zengzhong Li, Mao
  Yang, and Jingdong Wang.
\newblock {SPANN}: Highly-efficient billion-scale approximate nearest neighbor
  search.
\newblock In {\em Advances in Neural Information Processing Systems (NeurIPS)},
  volume~34, pages 5199--5212, 2021.

\bibitem{alayaLaser}
Weijian Chen, Haotian Liu, Yangshen Deng, Long Xiang, Liang Huang, and Bo~Tang.
\newblock Efficient index layout and search strategy for large-scale
  high-dimensional vector similarity search.
\newblock {\em Proceedings of the ACM on Management of Data (SIGMOD)},
  4(3):1--27, 2026.

\bibitem{covington2016youtube}
Paul Covington, Jay Adams, and Emre Sargin.
\newblock Deep neural networks for {YouTube} recommendations.
\newblock In {\em Proceedings of the 10th ACM Conference on Recommender Systems
  (RecSys)}, pages 191--198, 2016.

\bibitem{dade}
Liwei Deng, Penghao Chen, Ximu Zeng, Tianfu Wang, Yan Zhao, and Kai Zheng.
\newblock Efficient data-aware distance comparison operations for
  high-dimensional approximate nearest neighbor search.
\newblock {\em Proceedings of the VLDB Endowment (PVLDB)}, 18(3):812--821,
  2024.

\bibitem{faiss}
Matthijs Douze, Alexandr Guzhva, Chengqi Deng, Jeff Johnson, Gergely Szilvasy,
  Pierre-Emmanuel Mazar{\'e}, Maria Lomeli, Lucas Hosseini, and Herv{\'e}
  J{\'e}gou.
\newblock The {Faiss} library.
\newblock {\em arXiv preprint arXiv:2401.08281}, 2024.

\bibitem{cloudlab}
Dmitry Duplyakin, Robert Ricci, Aleksander Maricq, Gary Wong, Jonathon Duerig,
  Eric Eide, Leigh Stoller, Mike Hibler, David Johnson, Kirk Webb, Aditya
  Akella, Kuangching Wang, Glenn Ricart, Larry Landweber, Chip Elliott, Michael
  Zink, Emmanuel Cecchet, Snigdha Kar, and Prabodh Mishra.
\newblock The design and operation of {CloudLab}.
\newblock In {\em Proceedings of the 2019 USENIX Annual Technical Conference
  (USENIX ATC)}, pages 1--14, 2019.

\bibitem{rabitq-ext}
Jianyang Gao, Yutong Gou, Yuexuan Xu, Yongyi Yang, Cheng Long, and Raymond
  Chi-Wing Wong.
\newblock Practical and asymptotically optimal quantization of high-dimensional
  vectors in {E}uclidean space for approximate nearest neighbor search.
\newblock {\em Proceedings of the ACM on Management of Data (SIGMOD)},
  3(3):1--26, 2025.

\bibitem{adsampling}
Jianyang Gao and Cheng Long.
\newblock High-dimensional approximate nearest neighbor search: with reliable
  and efficient distance comparison operations.
\newblock {\em Proceedings of the ACM on Management of Data (SIGMOD)},
  1(2):1--27, 2023.

\bibitem{rabitq}
Jianyang Gao and Cheng Long.
\newblock {RaBitQ}: Quantizing high-dimensional vectors with a theoretical
  error bound for approximate nearest neighbor search.
\newblock {\em Proceedings of the ACM on Management of Data (SIGMOD)},
  2(3):1--27, 2024.

\bibitem{filtered-diskann}
Siddharth Gollapudi, Neel Karia, Varun Sivashankar, Ravishankar Krishnaswamy,
  Nikit Begwani, Swapnil Raz, Yiyong Lin, Yin Zhang, Neelam Mahapatro,
  Premkumar Srinivasan, Amit Singh, and Harsha~Vardhan Simhadri.
\newblock {Filtered-DiskANN}: Graph algorithms for approximate nearest neighbor
  search with filters.
\newblock In {\em Proceedings of the ACM Web Conference 2023 (WWW)}, pages
  3406--3416, 2023.

\bibitem{symphonyqg}
Yutong Gou, Jianyang Gao, Yuexuan Xu, and Cheng Long.
\newblock {SymphonyQG}: Towards symphonious integration of quantization and
  graph for approximate nearest neighbor search.
\newblock {\em Proceedings of the ACM on Management of Data (SIGMOD)},
  3(1):1--26, 2025.

\bibitem{pipeann}
Hao Guo and Youyou Lu.
\newblock Achieving low-latency graph-based vector search via aligning
  best-first search algorithm with {SSD}.
\newblock In {\em Proceedings of the 19th USENIX Symposium on Operating Systems
  Design and Implementation (OSDI)}, pages 171--186, 2025.

\bibitem{scann}
Ruiqi Guo, Philip Sun, Erik Lindgren, Quan Geng, David Simcha, Felix Chern, and
  Sanjiv Kumar.
\newblock Accelerating large-scale inference with anisotropic vector
  quantization.
\newblock In {\em Proceedings of the 37th International Conference on Machine
  Learning (ICML)}, pages 3887--3896, 2020.

\bibitem{anthropic-multiagent}
Jeremy Hadfield, Barry Zhang, Kenneth Lien, Florian Scholz, Jeremy Fox, and
  Daniel Ford.
\newblock How we built our multi-agent research system.
\newblock Anthropic Engineering Blog,
  \url{https://www.anthropic.com/engineering/multi-agent-research-system},
  2025.

\bibitem{clerc}
Abe~Bohan Hou, Orion Weller, Guanghui Qin, Eugene Yang, Dawn Lawrie, Nils
  Holzenberger, Andrew Blair-Stanek, and Benjamin Van~Durme.
\newblock {CLERC}: A dataset for {U}. {S}. legal case retrieval and
  retrieval-augmented analysis generation.
\newblock In {\em Findings of the Association for Computational Linguistics:
  NAACL 2025}, pages 7913--7928, 2025.

\bibitem{amx}
Intel Corporation.
\newblock {\em {Intel} Architecture Instruction Set Extensions and Future
  Features Programming Reference}, 2024.
\newblock Document Number 319433-053, June 2024; Chapter 3 covers the Intel
  {AMX} tile and {TMUL} architecture.

\bibitem{pq}
Herv{\'e} J{\'e}gou, Matthijs Douze, and Cordelia Schmid.
\newblock Product quantization for nearest neighbor search.
\newblock {\em IEEE Transactions on Pattern Analysis and Machine Intelligence
  (TPAMI)}, 33(1):117--128, 2011.

\bibitem{gustann}
Haodi Jiang, Hao Guo, Minhui Xie, Jiwu Shu, and Youyou Lu.
\newblock High-throughput, cost-effective billion-scale vector search with a
  single {GPU}.
\newblock {\em Proceedings of the ACM on Management of Data (SIGMOD)},
  3(6):1--27, 2025.

\bibitem{rago}
Wenqi Jiang, Suvinay Subramanian, Cat Graves, Gustavo Alonso, Amir
  Yazdanbakhsh, and Vidushi Dadu.
\newblock {RAGO}: Systematic performance optimization for retrieval-augmented
  generation serving.
\newblock In {\em Proceedings of the 52nd Annual International Symposium on
  Computer Architecture (ISCA)}, pages 974--989, 2025.

\bibitem{piperag}
Wenqi Jiang, Shuai Zhang, Boran Han, Jie Wang, Bernie Wang, and Tim Kraska.
\newblock {PipeRAG}: Fast retrieval-augmented generation via adaptive pipeline
  parallelism.
\newblock In {\em Proceedings of the 31st ACM SIGKDD Conference on Knowledge
  Discovery and Data Mining (KDD)}, pages 589--600, 2025.

\bibitem{searchr1}
Bowen Jin, Hansi Zeng, Zhenrui Yue, Jinsung Yoon, Sercan~O. Arik, Dong Wang,
  Hamed Zamani, and Jiawei Han.
\newblock {Search-R1}: Training {LLMs} to reason and leverage search engines
  with reinforcement learning.
\newblock In {\em Proceedings of the 2nd Conference on Language Modeling
  (COLM)}, 2025.

\bibitem{flashrag}
Jiajie Jin, Yutao Zhu, Zhicheng Dou, Guanting Dong, Xinyu Yang, Chenghao Zhang,
  Tong Zhao, Zhao Yang, and Ji-Rong Wen.
\newblock {FlashRAG}: A modular toolkit for efficient retrieval-augmented
  generation research.
\newblock In {\em Companion Proceedings of the ACM Web Conference 2025 (WWW)},
  pages 737--740, 2025.

\bibitem{curator}
Yicheng Jin, Yongji Wu, Wenjun Hu, Bruce~M. Maggs, Jun Yang, Xiao Zhang, and
  Danyang Zhuo.
\newblock Curator: Efficient vector search with low-selectivity filters.
\newblock {\em Proceedings of the ACM on Management of Data (SIGMOD)},
  4(1):1--27, 2026.

\bibitem{pgvector}
Andrew Kane.
\newblock {pgvector}: Open-source vector similarity search for {Postgres}.
\newblock \url{https://github.com/pgvector/pgvector}.

\bibitem{pageann}
Dingyi Kang, Dongming Jiang, Hanshen Yang, Hang Liu, and Bingzhe Li.
\newblock Scalable disk-based approximate nearest neighbor search with
  page-aligned graph.
\newblock {\em arXiv preprint arXiv:2509.25487}, 2025.

\bibitem{bang}
{Karthik V.}, Saim Khan, Somesh Singh, Harsha~Vardhan Simhadri, and Jyothi
  Vedurada.
\newblock {BANG}: Billion-scale approximate nearest neighbor search using a
  single {GPU}.
\newblock {\em arXiv preprint arXiv:2401.11324}, 2024.

\bibitem{cabana}
Minho Kim, Houxiang Ji, Jaeyoung Kang, Hwanjun Lee, Daehoon Kim, and Nam~Sung
  Kim.
\newblock {CABANA}: Cluster-aware query batching for accelerating billion-scale
  {ANNS} with {I}ntel {AMX}.
\newblock {\em IEEE Computer Architecture Letters (CAL)}, 24(2):289--292, 2025.

\bibitem{bioasq}
Anastasia Krithara, Anastasios Nentidis, Konstantinos Bougiatiotis, and
  Georgios Paliouras.
\newblock {BioASQ-QA}: A manually curated corpus for biomedical question
  answering.
\newblock {\em Scientific Data}, 10:170, 2023.

\bibitem{kuffo2025pdx}
Leonardo Kuffo, Elena Krippner, and Peter Boncz.
\newblock {PDX}: A data layout for vector similarity search.
\newblock {\em Proceedings of the ACM on Management of Data (SIGMOD)},
  3(3):1--26, 2025.

\bibitem{vllm}
Woosuk Kwon, Zhuohan Li, Siyuan Zhuang, Ying Sheng, Lianmin Zheng, Cody~Hao Yu,
  Joseph Gonzalez, Hao Zhang, and Ion Stoica.
\newblock Efficient memory management for large language model serving with
  {PagedAttention}.
\newblock In {\em Proceedings of the 29th ACM Symposium on Operating Systems
  Principles (SOSP)}, pages 611--626, 2023.

\bibitem{nemoann}
Nakyung Lee, Soobin Cho, Jiwoong Park, and Gyuyeong Kim.
\newblock {NemoANN}: {I/O}-efficient filtered vector search on {SSDs} with
  graph tunneling.
\newblock In {\em Proceedings of the ACM Symposium on Cloud Computing (SoCC)},
  2026.

\bibitem{lewis2020rag}
Patrick Lewis, Ethan Perez, Aleksandra Piktus, Fabio Petroni, Vladimir
  Karpukhin, Naman Goyal, Heinrich K{\"u}ttler, Mike Lewis, Wen-tau Yih, Tim
  Rockt{\"a}schel, Sebastian Riedel, and Douwe Kiela.
\newblock Retrieval-augmented generation for knowledge-intensive {NLP} tasks.
\newblock In {\em Advances in Neural Information Processing Systems (NeurIPS)},
  volume~33, pages 9459--9474, 2020.

\bibitem{saq}
Hui Li, Shiyuan Deng, Xiao Yan, Xiangyu Zhi, and James Cheng.
\newblock {SAQ}: Pushing the limits of vector quantization through code
  adjustment and dimension segmentation.
\newblock {\em Proceedings of the ACM on Management of Data (SIGMOD)},
  3(6):1--25, 2025.

\bibitem{sieve-ann}
Zhaoheng Li, Silu Huang, Wei Ding, Yongjoo Park, and Jianjun Chen.
\newblock {SIEVE}: Effective filtered vector search with collection of indexes.
\newblock {\em Proceedings of the VLDB Endowment (PVLDB)}, 18(11):4723--4736,
  2025.

\bibitem{postgres-fvs}
Duo Lu, Helena Caminal, Manos Chatzakis, Yannis Papakonstantinou, Yannis
  Chronis, Vaibhav Jain, and Fatma {\"O}zcan.
\newblock {An In-Depth Study of Filter-Agnostic Vector Search on a PostgreSQL
  Database System: [Experiments \& Analysis]}.
\newblock {\em Proceedings of the ACM on Management of Data}, 4(3), 2026.

\bibitem{hnsw}
Yury~A. Malkov and Dmitry~A. Yashunin.
\newblock Efficient and robust approximate nearest neighbor search using
  hierarchical navigable small world graphs.
\newblock {\em IEEE Transactions on Pattern Analysis and Machine Intelligence
  (TPAMI)}, 42(4):824--836, 2020.

\bibitem{quake}
Jason Mohoney, Devesh Sarda, Mengze Tang, Shihabur~Rahman Chowdhury, Anil
  Pacaci, Ihab~F. Ilyas, Theodoros Rekatsinas, and Shivaram Venkataraman.
\newblock Quake: Adaptive indexing for vector search.
\newblock In {\em Proceedings of the 19th USENIX Symposium on Operating Systems
  Design and Implementation (OSDI)}, pages 153--169, 2025.

\bibitem{naumov2019dlrm}
Maxim Naumov, Dheevatsa Mudigere, Hao-Jun~Michael Shi, Jianyu Huang, Narayanan
  Sundaraman, Jongsoo Park, Xiaodong Wang, Udit Gupta, Carole-Jean Wu,
  Alisson~G. Azzolini, Dmytro Dzhulgakov, Andrey Mallevich, Ilia Cherniavskii,
  Yinghai Lu, Raghuraman Krishnamoorthi, Ansha Yu, Volodymyr Kondratenko,
  Stephanie Pereira, Xianjie Chen, Wenlin Chen, Vijay Rao, Bill Jia, Liang
  Xiong, and Misha Smelyanskiy.
\newblock Deep learning recommendation model for personalization and
  recommendation systems.
\newblock {\em arXiv preprint arXiv:1906.00091}, 2019.

\bibitem{cagra}
Hiroyuki Ootomo, Akira Naruse, Corey Nolet, Ray Wang, Tamas Feher, and Yong
  Wang.
\newblock {CAGRA}: Highly parallel graph construction and approximate nearest
  neighbor search for {GPUs}.
\newblock In {\em Proceedings of the 40th IEEE International Conference on Data
  Engineering (ICDE)}, pages 4236--4247, 2024.

\bibitem{acorn}
Liana Patel, Peter Kraft, Carlos Guestrin, and Matei Zaharia.
\newblock {ACORN}: Performant and predicate-agnostic search over vector
  embeddings and structured data.
\newblock {\em Proceedings of the ACM on Management of Data (SIGMOD)},
  2(3):1--27, 2024.

\bibitem{pubmedmv}
{Qdrant}.
\newblock {PubMed-MV}: {BGE-M3} embeddings of {PubMed} abstracts.
\newblock \url{https://huggingface.co/datasets/Qdrant/PubMed-MV}, 2025.

\bibitem{qwen3}
{Qwen Team}.
\newblock {Qwen3} technical report.
\newblock {\em arXiv preprint arXiv:2505.09388}, 2025.

\bibitem{navix}
Gaurav Sehgal and Semih Saliho{\u{g}}lu.
\newblock {NaviX}: A native vector index design for graph {DBMSs} with robust
  predicate-agnostic search performance.
\newblock {\em Proceedings of the VLDB Endowment (PVLDB)}, 18(11):4438--4450,
  2025.

\bibitem{bigann21}
Harsha~Vardhan Simhadri, George Williams, Martin Aum{\"u}ller, Matthijs Douze,
  Artem Babenko, Dmitry Baranchuk, Qi~Chen, Lucas Hosseini, Ravishankar
  Krishnaswamy, Gopal Srinivasa, Suhas~Jayaram Subramanya, and Jingdong Wang.
\newblock Results of the {NeurIPS}'21 challenge on billion-scale approximate
  nearest neighbor search.
\newblock In {\em Proceedings of the NeurIPS 2021 Competitions and
  Demonstrations Track}, volume 176 of {\em Proceedings of Machine Learning
  Research}, pages 177--189, 2022.

\bibitem{trim}
Yitong Song, Pengcheng Zhang, Chao Gao, Bin Yao, Kai Wang, Zongyuan Wu, and Lin
  Qu.
\newblock {TRIM}: Accelerating high-dimensional vector similarity search with
  enhanced triangle-inequality-based pruning.
\newblock {\em Proceedings of the ACM on Management of Data (SIGMOD)},
  3(6):1--26, 2025.

\bibitem{diskann}
Suhas~Jayaram Subramanya, Devvrit, Rohan Kadekodi, Ravishankar Krishnaswamy,
  and Harsha~Vardhan Simhadri.
\newblock {DiskANN}: Fast accurate billion-point nearest neighbor search on a
  single node.
\newblock In {\em Advances in Neural Information Processing Systems (NeurIPS)},
  volume~32, pages 13748--13758, 2019.

\bibitem{leanvec}
Mariano Tepper, Ishwar~Singh Bhati, Cecilia Aguerrebere, Mark Hildebrand, and
  Theodore~L. Willke.
\newblock {LeanVec}: Searching vectors faster by making them fit.
\newblock {\em Transactions on Machine Learning Research (TMLR)}, 2024.

\bibitem{gleanvec}
Mariano Tepper, Ishwar~Singh Bhati, Cecilia Aguerrebere, and Ted Willke.
\newblock {GleanVec}: Accelerating vector search with minimalist nonlinear
  dimensionality reduction, 2024.

\bibitem{beir}
Nandan Thakur, Nils Reimers, Andreas R{\"u}ckl{\'e}, Abhishek Srivastava, and
  Iryna Gurevych.
\newblock {BEIR}: A heterogeneous benchmark for zero-shot evaluation of
  information retrieval models.
\newblock In {\em Proceedings of the Neural Information Processing Systems
  Track on Datasets and Benchmarks}, 2021.

\bibitem{smartanns}
Bing Tian, Haikun Liu, Zhuohui Duan, Xiaofei Liao, Hai Jin, and Yu~Zhang.
\newblock Scalable billion-point approximate nearest neighbor search using
  {SmartSSDs}.
\newblock In {\em Proceedings of the 2024 USENIX Annual Technical Conference
  (ATC)}, pages 1135--1150, 2024.

\bibitem{fusionanns}
Bing Tian, Haikun Liu, Yuhang Tang, Shihai Xiao, Zhuohui Duan, Xiaofei Liao,
  Hai Jin, Xuecang Zhang, Junhua Zhu, and Yu~Zhang.
\newblock Towards high-throughput and low-latency billion-scale vector search
  via {CPU/GPU} collaborative filtering and re-ranking.
\newblock In {\em Proceedings of the 23rd USENIX Conference on File and Storage
  Technologies (FAST)}, pages 171--185, 2025.

\bibitem{ircot}
Harsh Trivedi, Niranjan Balasubramanian, Tushar Khot, and Ashish Sabharwal.
\newblock Interleaving retrieval with chain-of-thought reasoning for
  knowledge-intensive multi-step questions.
\newblock In {\em Proceedings of the 61st Annual Meeting of the Association for
  Computational Linguistics (ACL)}, pages 10014--10037, 2023.

\bibitem{milvus}
Jianguo Wang, Xiaomeng Yi, Rentong Guo, Hai Jin, Peng Xu, Shengjun Li, Xiangyu
  Wang, Xiangzhou Guo, Chengming Li, Xiaohai Xu, Kun Yu, Yuxing Yuan, Yinghao
  Zou, Jiquan Long, Yudong Cai, Zhenxiang Li, Zhifeng Zhang, Yihua Mo, Jun Gu,
  Ruiyi Jiang, Yi~Wei, and Charles Xie.
\newblock Milvus: A purpose-built vector data management system.
\newblock In {\em Proceedings of the 2021 ACM SIGMOD International Conference
  on Management of Data (SIGMOD)}, pages 2614--2627, 2021.

\bibitem{nhq}
Mengzhao Wang, Lingwei Lv, Xiaoliang Xu, Yuxiang Wang, Qiang Yue, and Jiongkang
  Ni.
\newblock An efficient and robust framework for approximate nearest neighbor
  search with attribute constraint.
\newblock In {\em Advances in Neural Information Processing Systems (NeurIPS)},
  volume~36, 2023.

\bibitem{starling}
Mengzhao Wang, Weizhi Xu, Xiaomeng Yi, Songlin Wu, Zhangyang Peng, Xiangyu Ke,
  Yunjun Gao, Xiaoliang Xu, Rentong Guo, and Charles Xie.
\newblock Starling: An {I/O}-efficient disk-resident graph index framework for
  high-dimensional vector similarity search on data segment.
\newblock {\em Proceedings of the ACM on Management of Data (SIGMOD)},
  2(1):1--27, 2024.

\bibitem{vecflow}
Jingyi Xi, Chenghao Mo, Ben Karsin, Artem Chirkin, Mingqin Li, and Minjia
  Zhang.
\newblock {VecFlow: A High-Performance Vector Data Management System for
  Filtered-Search on GPUs}.
\newblock {\em Proceedings of the ACM on Management of Data}, 3(4), 2025.

\bibitem{flashanns}
Yang Xiao, Mo~Sun, Ziyu Song, Bing Tian, Jie Sun, Jie Zhang, Zeke Wang, Zonghui
  Wang, Wenzhi Chen, and Fei Wu.
\newblock {FlashANNS}: {GPU}-driven asynchronous {I/O} pipelining for
  eliminating storage-compute bottlenecks in billion-scale similarity search.
\newblock {\em Proceedings of the ACM on Management of Data (SIGMOD)},
  4(1):1--27, 2026.

\bibitem{medrag}
Guangzhi Xiong, Qiao Jin, Zhiyong Lu, and Aidong Zhang.
\newblock Benchmarking retrieval-augmented generation for medicine.
\newblock In {\em Findings of the Association for Computational Linguistics:
  ACL 2024}, pages 6233--6251, 2024.

\bibitem{harmony}
Qian Xu, Feng Zhang, Chengxi Li, Lei Cao, Zheng Chen, Jidong Zhai, and Xiaoyong
  Du.
\newblock {HARMONY: A Scalable Distributed Vector Database for High-Throughput
  Approximate Nearest Neighbor Search}.
\newblock {\em Proceedings of the ACM on Management of Data}, 3(4):1--28, 2025.

\bibitem{rairs}
Zehai Yang and Shimin Chen.
\newblock {RAIRS}: Optimizing redundant assignment and list layout for
  {IVF}-based {ANN} search.
\newblock {\em Proceedings of the ACM on Management of Data (SIGMOD)},
  4(1):1--27, 2026.

\bibitem{react}
Shunyu Yao, Jeffrey Zhao, Dian Yu, Nan Du, Izhak Shafran, Karthik Narasimhan,
  and Yuan Cao.
\newblock {ReAct}: Synergizing reasoning and acting in language models.
\newblock In {\em Proceedings of the 11th International Conference on Learning
  Representations (ICLR)}, 2023.

\bibitem{margo}
Ziyang Yue, Bolong Zheng, Ling Xu, Kanru Xu, Shuhao Zhang, Yajuan Du, Yunjun
  Gao, Xiaofang Zhou, and Christian~S. Jensen.
\newblock Select edges wisely: Monotonic path aware graph layout optimization
  for disk-based {ANN} search.
\newblock {\em Proceedings of the VLDB Endowment (PVLDB)}, 18(11):4337--4349,
  2025.

\bibitem{adaef}
Chao Zhang and Ren{\'e}e~J. Miller.
\newblock {Distribution-Aware Exploration for Adaptive HNSW Search}.
\newblock {\em Proceedings of the ACM on Management of Data}, 4(1):1--27, 2026.

\bibitem{vbase}
Qianxi Zhang, Shuotao Xu, Qi~Chen, Guoxin Sui, Jiadong Xie, Zhizhen Cai, Yaoqi
  Chen, Yinxuan He, Yuqing Yang, Fan Yang, Mao Yang, and Lidong Zhou.
\newblock {VBASE}: Unifying online vector similarity search and relational
  queries via relaxed monotonicity.
\newblock In {\em Proceedings of the 17th USENIX Symposium on Operating Systems
  Design and Implementation (OSDI)}, pages 377--395, 2023.

\bibitem{housingqa}
Lucia Zheng, Neel Guha, Javokhir Arifov, Sarah Zhang, Michal Skreta,
  Christopher~D. Manning, Peter Henderson, and Daniel~E. Ho.
\newblock A reasoning-focused legal retrieval benchmark.
\newblock In {\em Proceedings of the Symposium on Computer Science and Law
  (CSLAW)}, 2025.

\bibitem{cotra}
Xiangyu Zhi, Meng Chen, Xiao Yan, Baotong Lu, Hui Li, Qianxi Zhang, Qi~Chen,
  and James Cheng.
\newblock {CoTra: Towards Efficient and Scalable Distributed Vector Search with
  RDMA}.
\newblock {\em Proceedings of the ACM on Management of Data}, 4(1), 2026.

\bibitem{vdbbench}
{Zilliz}.
\newblock {VectorDBBench}: A benchmark tool for vector databases.
\newblock \url{https://github.com/zilliztech/VectorDBBench}, 2023.

\end{thebibliography}

\end{document}